\documentclass[fleqn,usenatbib]{mnras}

\usepackage{newtxtext,newtxmath}
\usepackage[T1]{fontenc}

\DeclareRobustCommand{\VAN}[3]{#2}
\let\VANthebibliography\thebibliography
\def\thebibliography{\DeclareRobustCommand{\VAN}[3]{##3}\VANthebibliography}

\usepackage{graphicx}	% Including figure files
\usepackage{amsmath}	% Advanced maths commands
\title[Superbubbles and Winds in FIRE-2]{Any Way the Wind Blows: Quantifying Superbubbles and their Outflows in Simulated Galaxies across $z \approx 0-3$}

\author[L. E. Porter et al.]{
Lori E. Porter,$^{1}$\thanks{E-mail: lep2176@columbia.edu}
Matthew E. Orr,$^{2,3}$
Blakesley Burkhart,$^{2,3}$
Andrew Wetzel,$^{4}$  
Dušan Kereš,$^{5}$ \newauthor
Claude-André Faucher-Giguère,$^{6}$ 
\& Philip F. Hopkins$^{7}$ 
\\
$^{1}$Department of Astronomy, Columbia University, 538 West 120 Street, New York, NY 10027, USA\\
$^{2}$Center for Computational Astrophysics, Flatiron Institute, 162 Fifth Avenue, New York, NY 10010, USA\\
$^{3}$Department of Physics and Astronomy, Rutgers University, 136 Frelinghuysen Road, Piscataway, NJ 08854, USA\\
$^{4}$ Department of Physics and Astronomy, University of California, Davis, CA 95616, USA \\
$^{5}$Department of Physics, Center for Astrophysics and Space Science, University of California at San Diego, 9500 Gilman Drive, La Jolla, CA 92093, USA\\
$^{6}$Department of Physics and Astronomy and CIERA, Northwestern University, 2145 Sheridan Road, Evanston, IL 60208, USA\\
$^{7}$TAPIR, California Institute of Technology, 1200 E. California Blvd., MC 350-17, Pasadena, CA 91125, USA\\
}

\date{Accepted XXX. Received YYY; in original form ZZZ}

\pubyear{2024}

\begin{document}
\label{firstpage}
\pagerange{\pageref{firstpage}--\pageref{lastpage}}
\maketitle

\begin{abstract}
We present an investigation of clustered stellar feedback in the form of superbubbles identified within eleven galaxies from the FIRE-2 (Feedback in Realistic Environments) cosmological zoom-in simulation suite, at both cosmic noon (1 < z < 3) and in the local Universe. We study the spatially-resolved multiphase outflows that these supernovae drive, comparing our findings with recent theory and observations. These simulations consist of five LMC-mass galaxies and six Milky Way-mass progenitors (with a minimum baryonic particle mass of $m_{b.min} = 7100 M_{\odot}$), for which we calculate the local mass and energy loading factors on 750~pc scales from the identified outflows. We also characterize the multiphase morphology and properties of the identified superbubbles, including the `shell' of cool ($T<10^5$ K) gas and break out of energetic hot ($T>10^5$ K) gas when the shell bursts. For all galaxies, the outflow mass, momentum, and energy fluxes appear to reach their peak during the identified superbubbles, and we investigate the effects on the interstellar medium (ISM), circumgalactic medium (CGM), and subsequent star formation rates. We find that these simulations, regardless of redshift, have mass-loading factors and momentum fluxes in the cool gas that largely agree with recent observations. Lastly, we also investigate how methodological choices in measuring outflows can affect loading factors for galactic winds.
\end{abstract}


%%%%%%%%%%%%%%%%%%%%%%%%%%%%%%%%%%%%%%%%%%%%%%%%%%

%%%%%%%%%%%%%%%%% BODY OF PAPER %%%%%%%%%%%%%%%%%%

\section{Introduction}
\label{sec:intro}

Stellar feedback is one of the most important factors driving galaxy evolution. This feedback plays a critical role in the structure of the interstellar medium (ISM), as supernova (SN) explosions are a primary source of momentum, energy, and mass injection back into the ISM.  Therefore, stellar feedback must be properly understood in order to accurately reproduce observed relationships and properties of galaxies, including the mass-metallicity relation \citep{Tremonti2004, Dave2011, Ma2016, Wetzel2016, Porter2022, Muratov2017, Bassini2024, Marszewski2024}, turbulence \citep{Faucher-Giguere2013, Orr2018, Orr2020, Burkhart2021}, low star formation efficiencies \citep{Hopkins2014,Burkhart2018}, stellar masses \citep{Dave2012, Wetzel2016, Agertz2016}, and the baryon cycle \citep{Muratov2015, Angles-Alcazar2017a, Kim2017b, Shin2023}.

However, star formation and the resultant feedback do not occur uniformly within a galaxy. Instead, it occurs in clusters \citep{Motte2018, Krumholz2019, Tacconi2020}, and indeed \citet{Fielding2018} and \citet{Governato2010} note the necessity of such clustering in driving galactic winds. As a consequence of the proximity of formation of stars within a cluster, when core-collapse SNe from massive stars occur their shock fronts coalesce, and an even larger encompassing shock front known as a superbubble is born.

Superbubbles can be found in galaxies both at high-redshift (z $\geq$ 1) and in the local Universe \citep{Taniguchi2001, Keller2015, Orr2022a, Watkins2023a}. They typically consist of a hot, volume-filling component that sweeps up the surrounding ISM as it expands, creating a "hole" in which very little cold gas resides. As the bubbles travel through the ISM, they capture cold molecular clouds that become entrained or homogenized within the hot and diffuse winds. This phenomenon results in the shell-like morphology of cold gas,  referred to as the `cold cap' of the superbubble in \citet{Orr2022}, around the energetic hot gas interior with evaporating cloudlet inclusions \citep{Lancaster2021}. 

As a result of their importance, superbubbles have been widely studied for decades \citep{Castor1975, Bregman1980, MacLow1988, MacLow1989, Ostriker1988, Koo1992}, but it is with recent computational methods and observational facilities that we truly become able to resolve the physics of superbubbles and their relationship with the baryon cycle and galaxy evolution. Notably, work enabled by JWST has begun to quantify the effects of superbubbles in nearby galaxies, such as NGC 628, in driving turbulence, sweeping up dense gas, and triggering additional star formation along the expanding shock fronts \citep{Barnes2023, Mayya2023, Watkins2023a, Watkins2023b}. 

The disturbance of the ISM by superbubbles intimately connects galactic outflows and dense gas turbulence, as the successful breakout of bubbles from the galactic disk can drive some of the largest winds in galaxies \citep{Muratov2015, Kim2017a, Fielding2018, Martizzi2020}. These multiphase winds, integral to the Baryon cycle, are essential to the process of galaxy evolution \citep{Tumlinson2017, Faucher-Giguere2023}. Inflowing winds (meaning, reaccreting wind material) carry material necessary to sustain star formation and black hole growth, such as cold molecular gas, while new outflows from breakouts transport material into the circumgalactic medium (CGM). As a result, these winds (and the stellar feedback that drives them) are key to regulating star formation in galaxies \citep{Oppenheimer2010, Ostriker2010, Hopkins2014, Hayward2017}. 

Outflow loading factors, often established in terms of mass \citep[$\eta_M$;][]{Larson1974b, Veilleux2005} and energy \citep[$\eta_E$;][]{Larson1974b, Chevalier1985, Strickland2009}, are used to quantify the properties of winds. These measures are evaluated as the respective outflowing quantity (e.g., mass or energy) normalized by the star formation rate, resulting in a dimensionless quantity that describes an efficiency relative to star formation. 
Loading factors have been frequently used in simulations to investigate and quantify galactic outflows and galaxy evolution \citep{Muratov2015, Kim2020a, Mitchell2020, Pandya2021, Steinwandel2024}, and while loading factors are more difficult to ascertain in observations, there still exists a sample to compare against theory \citep{Martin1999, Heckman2015, Chisholm2017, McQuinn2019, ReichardtChu2022, McPherson2024}. However, measurements can vary across several orders of magnitude based on galaxy properties and dynamics. Several studies find that mass loading ($\eta_M$) is dependent on current star formation surface density ($\Sigma_{\rm SFR}$) as $\eta_M \propto \Sigma_{\rm SFR}^{-0.44}$ \citep{Li2017, Kim2020a, Li2020}. Momentum-driven outflows, on the other hand, predict an even steeper dependence when outward velocity ($v_{out}$) is considered \citep[$v_{out} \propto \Sigma_{\rm SFR}^2$; $v_{out} \propto \rm SFR$;][]{Murray2011, Hopkins2012}{}{}. Despite the variations in the dependence of outflows and loading factors on current and recent star formation, they have a well-studied relationship and aid in explaining the role of galactic winds in LMC-mass galaxies \citep{Fielding2018, McQuinn2019, Romano2023}, and the mass-metallicity relation \citep{Finlator2008, Ma2016, Chisholm2018, Bassini2024}.

Simulations have become a powerful tool for studying superbubbles and the multiphase winds they can drive. However, the physical nature of these winds pose additional complications and restricts reliable studies to only simulations that can resolve a multiphase ISM. Another challenge for investigating superbubbles' effect on their simulated host galaxies is that there is no standard method for identifying and quantifying galactic outflows. For example, some groups quantify outflows through fixed surfaces above and below galaxies (not unlike observational work) and take cuts on velocity relative to local escape velocity or the Bernoulli velocity \citep{Kim2020a, Pandya2021}, while others use particle tracking methods to differentiate outflows from fountains \citep{Angles-Alcazar2017a, Hafen2019}.

\citet{Orr2022a} developed an analytic model of clustered feedback from SNe, finding that the local gas fraction and dynamical time determine whether superbubbles broke out of the ISM (driving winds) or fragmented within the galaxy (driving turbulence). \citet{Orr2022} compared this theory work with local observations from \citet{Barnes2023}, \citet{Mayya2023}, and \citet{Watkins2023a}, finding general agreement. The work presented in this paper aims, in part, to compare these predictions to simulations in local galaxies and galaxies at cosmic noon.

In particular, this paper makes use of eleven galaxies from the FIRE-2 (Feedback in Realistic Environments\footnote{\url{http://fire.northwestern.edu}}; \citealt{Hopkins2018}) suite of cosmological zoom-in simulations, identifying and quantifying outflows from within these galaxies (five low-mass and six Milky Way-mass) at both high redshift (z $\sim$ 1-3) and the local Universe (z $\sim$ 0). These galaxies and cosmological epochs represent a diverse range over which we can study the launching of galactic winds and their properties, including superbubbles, as well as compare with recent observations.

We begin in Section~\ref{sec:meth} by detailing the FIRE-2 simulations and how we choose to define outflows within them. In Section~\ref{sec:results} we present our results, including the wind properties and corresponding loading factors (Section~\ref{subsec:overallfluxes}), and their connection to superbubbles (Section~\ref{subsec:superbubbles}). The physical implications of these results are further discussed in comparison to other simulations and observations in Section~\ref{sec:disc}, and we briefly summarize our findings in Section~\ref{sec:conc}.

\section{Simulations \& Analysis Method}\label{sec:meth}

We investigate superbubble feedback events in six Milky Way/Andromeda-mass galaxies, and five Magellanic Cloud (SMC/LMC)-mass galaxies (these masses all being achieved at roughly $z\approx 0$), from the `standard physics' FIRE-2 simulations introduced in \citet{Hopkins2018}, which are publicly available \citep{Wetzel2023}.
This work makes use of $\sim$10 snapshots of each galaxy near each integer redshift from zero to three (for a total of $\sim$40 snapshots per galaxy over its evolution), with an approximate time spacing between snapshots of $\sim$25 Myr. Table~\ref{table:gal_props} briefly summarizes the global gas and stellar mass properties of the simulations analyzed here at each integer redshift.  

The simulations analyzed in this paper have baryonic particle masses on the order of $m_{\rm b,min} = 7100$ M$_\odot${\footnote{Several of the low-mass m11 galaxies analyzed here are simulated at higher resolutions, but we choose to use the $m_{\rm b,min} = 7100$ M$_\odot$ simulations for consistency in our comparisons.}} and minimum adaptive force softening lengths $<$1~pc. Cooling in FIRE-2 is computed for gas temperatures $T=10-10^{10} K$. The suite of simulations includes a wide variety of heating and cooling physics, including free-free, photo-ionization/recombination, Compton, photo-electric, metal-line, molecular, fine-structure, and dust collisional processes. In particular, metal-line cooling is noted by \citet{Hopkins2018} to be particularly important for superbubbles.
The gas softening lengths are adaptive and we note that the effective radius of gas elements at the the minimum density of star formation is about 7 pc ($n=1000 \; \rm cm^{-3}$; see Table 3 of \citealt{Hopkins2018}), with the densest structures having shorter smoothing lengths down to the sub-pc minimum. 

In the FIRE-2 simulations, stars form on a free-fall time in gas that is dense ($n > 10^3$ cm$^{-3}$), molecular (according to the \citealt{Krumholz2011} methodology), self-gravitating (viral parameter $\alpha_{\rm vir} < 1$), and Jeans-unstable. Star particles are considered single stellar populations with defined age, metallicity, and mass. 

The FIRE-2 "standard physics" incorporates feedback mechanisms from supernovae, stellar winds from OB/AGB stars, photoionization and photoelectric heating, and radiation pressure. This suite excludes AGN, cosmic rays, and additional MHD physics, although other studies within the broader FIRE-2 project have explored these "extended physics" elements (\citealt{Angles-Alcazar2017, Chan2019, Su2019a}).
For comprehensive details on the "standard" physics and their application, see \citet{Hopkins2018}. Of particular importance to this current study of superbubbles/SN-driven winds are the core-collapse and Type Ia supernova rates, derived from \textit{STARBURST99} (\citealt{Leitherer1999}) and \citet{Mannucci2006}, respectively.

We generate maps of the gas, stellar, and star formation rate properties from the snapshots using the same methods as \citet{Orr2018} and \citet{Orr2020}, projecting the galaxies face-on (or edge-on) using the angular momentum of the star particles within the stellar half-mass radius, and binning star particles and gas cells into square pixels with side-lengths (\emph{i.e.}, ``pixel sizes'' {$l_{\rm pix}$}) 750~pc. The maps are 30~kpc on a side, and integrate gas and stars within $\pm 15$ kpc of the galactic mid-plane. Cold gas structures are well-contained within a single one of these pixels, and the most diffuse hot gas cells are marginally resolved by these pixels down to densities of $\sim$10$^{-3}$~cm$^{-3}$. The length scale is comparable with the resolvable scale of JWST observations at $z\sim1$ \citep{Boker2022}.

We generate a proxy for observational measures of recent SFRs by calculating the 40 Myr-averaged SFR. We do this by summing the mass of star particles with ages less than 40 Myr, and correcting for mass loss from stellar winds and evolutionary effects using predictions from {\scriptsize STARBURST99} \citep{Leitherer1999}.  
This time interval was chosen for its \emph{approximate} correspondence with the timescales traced by continuum UV-derived SFRs \citep{Lee2009}.

To calculate the outflow mass, momentum, and energy fluxes in a snapshot, we project the gas cells onto planes a fixed height above/below the galaxy and then calculate the flux quantities through that surface in a Cartesian grid of 750~pc pixels using the following definitions:
\begin{equation}
    \dot M_{\rm pix} \equiv l^2_{\rm pix} \sum_i \rho_i v_{i,z} ,
\end{equation}
\begin{equation}
    \dot P_{\rm pix} \equiv l^2_{\rm pix} \sum_i \rho_i [v_{i,z}^2 + (\gamma -1)u_i] , 
\end{equation}
and
\begin{equation}
    \dot E_{\rm pix} \equiv l^2_{\rm pix} \sum_i \rho_i v_{i,z} (v_i^2+\gamma u_i)  ,    
\end{equation}
where $i$ is summing over all the gas cells whose kernel overlaps with the pixel surface, $\rho$ is the gas element density evaluated at the pixel surface, $v_z$ is the gas velocity normal to the pixel surface, $v^2$ is the square of all velocity components\footnote{We note that definitionally, other authors either include or omit an additional factor of $1/2$ here in the energy flux definition, see, e.g., \citet{Kim2020a} vs.~\citet{Steinwandel2024}.}, $\gamma =5/3$ is the adiabatic index (i.e., heat capacity ratio) for a monatomic gas, and $u$ is the specific internal energy of the gas element.  For high-redshift snapshots ($z = 1-3$) we select a height above/below the main galaxy body of 0.05$R_{\rm vir}$, and for the low-redshift snapshots ($z \approx 0$) we select a height above/below the galaxy of twice the gas scale height $2H$. We note that where outflows are defined can have an effect on the resultant loading factors. We briefly analyze how different choices for the height of the flux surface affect our results in Appendix~\ref{flux_surface_changes}.

\begin{table*}\caption{Summary of FIRE-2 galaxy properties, including stellar mass, gas mass, and gas fraction, across all redshift ranges used in this work.}\label{table:gal_props}
    \begin{tabular}{lccccccccccccc} 
\hline
&  \multicolumn{3}{c}{$z\approx 3$} & \multicolumn{3}{c}{$z\approx 2$} & \multicolumn{3}{c}{$z\approx 1$} & \multicolumn{3}{c}{$z\approx 0$} \\
\hline
Name & $\log(\frac{M_\star}{{\rm M_\odot}})$ & $\log(\frac{M_{\rm gas}}{{\rm M_\odot}})$ & $\rm f_{gas}$ 
& $\log(\frac{M_\star}{{\rm M_\odot}})$ & $\log(\frac{M_{\rm gas}}{{\rm M_\odot}})$ & $\rm f_{gas}$ 
& $\log(\frac{M_\star}{{\rm M_\odot}})$ & $\log(\frac{M_{\rm gas}}{{\rm M_\odot}})$ & $\rm f_{gas}$ & $\log(\frac{M_\star}{{\rm M_\odot}})$ & $\log(\frac{M_{\rm gas}}{{\rm M_\odot}})$ & $\rm f_{gas}$\\ 
\hline
\textbf{m11d} 
& 7.90 & 9.32 & 0.94 & 8.30 & 9.23 & 0.83 & 8.65 & 8.81 & 0.49 & 9.63 & 9.64 & 0.47 \\
\textbf{m11e}
& 7.58 & 9.19 & 0.94 & 8.10 & 9.42 & 0.92 & 8.78 & 9.20 & 0.69 & 9.14 & 9.30 & 0.59 \\
\textbf{m11h}
& 8.22 & 9.38 & 0.91 & 8.59 & 9.58 & 0.87 & 9.10 & 9.61 & 0.69 & 9.59 & 9.60 & 0.47 \\
\textbf{m11i}
& 6.90 & 8.86 & 0.96 & 7.46 & 9.01 & 0.94 & 7.97 & 9.20 & 0.88 & 9.00 & 9.19 & 0.60 \\
\textbf{m11q}
& 7.97 & 9.01 & 0.89 & 8.25 & 8.95 & 0.8 & 8.55 & 8.89 & 0.66 & 8.82 & 9.14 & 0.66 \\
\textbf{m12b} 
&  9.01 & 9.81 & 0.83 & 9.80 & 10.04 & 0.61 & 10.51 & 9.97 & 0.14 & 10.96 & 10.30 & 0.14 \\
\textbf{m12c} 
& 8.95 & 9.63 & 0.77 & 9.26 & 9.68 & 0.66 & 10.18 & 10.18 & 0.48 & 10.80 & 10.29 & 0.14 \\
\textbf{m12f} 
& 9.24 & 9.99 & 0.91 & 9.96 & 10.18 & 0.77 & 10.43 & 10.02 & 0.34 & 10.92 & 10.30 & 0.14 \\
\textbf{m12i} 
& 9.01 & 10.04 & 0.84 & 9.60 & 10.15 & 0.56 & 10.29 & 10.06 & 0.20 & 10.84 & 10.30 & 0.18 \\
\textbf{m12m} 
& 8.52 & 9.61 & 0.85 & 9.53 & 10.07 & 0.77 & 10.39 & 10.43 & 0.52 & 11.09 & 10.38 & 0.14 \\
\textbf{m12r}
& 9.12 & 9.71 & 0.74 & 9.43 & 9.64 & 0.57 & 9.66 & 9.49 & 0.34 & 10.26 & 9.97 & 0.33 \\
\hline
%\multicolumn{13}{l}{Note: all quantities measured within a 30~kpc cubic aperture.}\\
\multicolumn{13}{l}{Simulations here are introduced by \citet{Wetzel2016}, \citet{El-Badry2018}, \citet{Hopkins2018}, and \citet{Samuel2020}.} \\
    \end{tabular}
\end{table*}

\subsection{Characterizing the Individual Simulations}\label{subsec:characterization}

\begin{figure*}
	\includegraphics[width=1\textwidth]{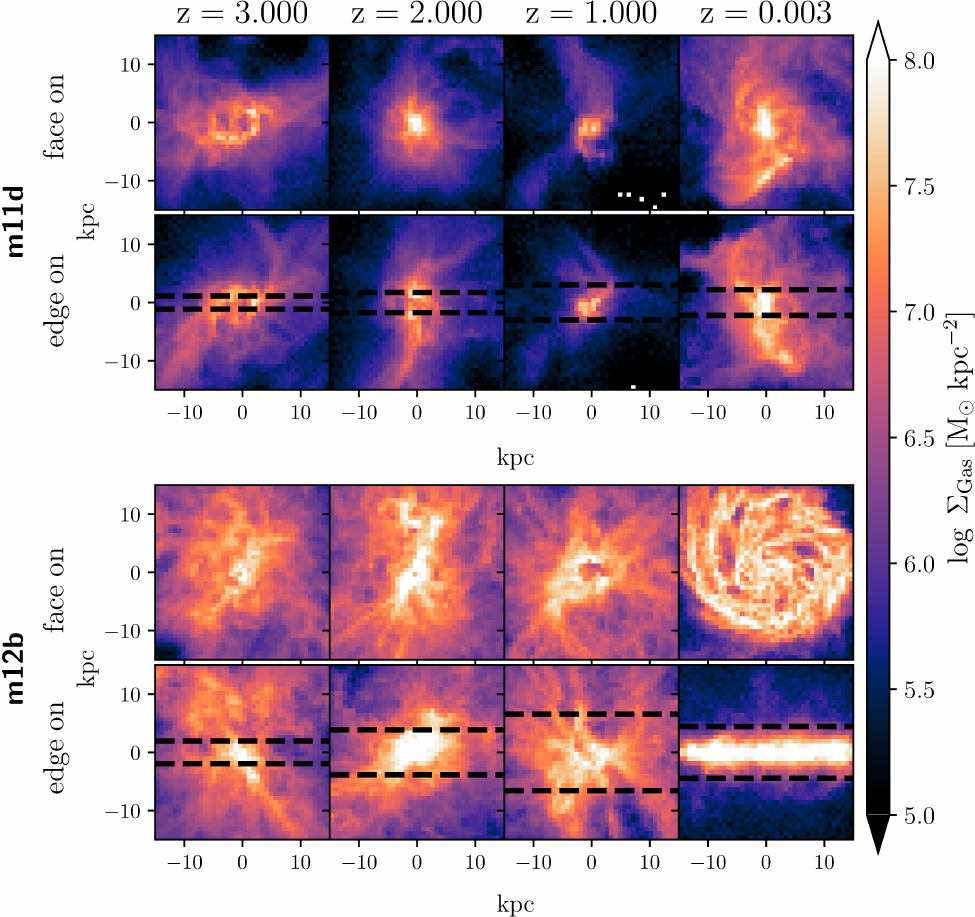}
    \caption{Face and edge-on spatial distributions of all gas in two galaxies at a representative snapshot in each redshift bin, with pixel size of 750pc. Top panels show evolution of a LMC-mass galaxy progenitor \textbf{m11d}, whereas bottom panels show that of a (disky at $z\approx 0 $) Milky Way-mass progenitor \textbf{m12b}.  Dashed black lines on edge-on panels represent the surfaces through which outflows are identified, located at 0.05~$R_{\rm vir}$ for $z=1-3$, and twice the galaxy average gas scale height ($2H$) for $z\approx 0$.}
    \label{fig:faceon_m11m12_gas_imshow}
\end{figure*}

We find it important to briefly review the galaxies' basic properties and behavior. While we use eleven FIRE-2 galaxies in this study, the majority of examples of specific superbubble events shown in the Figures will be from two galaxies for conciseness: one from the low-mass m11 galaxies, \textbf{m11d}, and one from the MW-mass galaxies, \textbf{m12b}. This allows us to showcase superbubbles and wind properties of both galaxy types, in addition to bubbles occurring at different redshifts. Table~\ref{table:gal_props} presents the stellar masses, gas masses, and gas fractions ($f_{\rm gas} = M_{\rm gas}/(M_{\rm gas}+M_{\rm \star})$ at the radius within which 90$\%$ of the galaxy snapshot's stellar mass is contained) of the simulated galaxies analyzed in this paper in each of the redshift bins from $z = 3-0$.

Our sample of low-mass galaxies (m11s) shows remarkable diversity in morphologies and modes of star formation for five galaxies of similar mass.  Only one LMC-mass galaxy, \textbf{m11h}, forms a disk, while the others are irregular galaxies with small starburst events.  They generally remain high in gas fraction ($f_{\rm gas} \gtrsim 0.5$) for their entire evolution down to $z =0$. The top panels of Figure~\ref{fig:faceon_m11m12_gas_imshow} illustrate the spatial distribution of all gas throughout galaxy \textbf{m11d}, with pixel size 750pc: this galaxy shows little distinct morphological features, though signs of strong stellar feedback are present at $z\sim3$. 

%As for the larger galaxies in our sample, the m12s experience a higher frequency of mergers, which is expected based on mergers that satisfy the $M_{*}-M_{\rm halo}$ relation \citep{Hopkins2018}. 
The larger galaxies in our sample (the m12s) all form disks by $z=0$ and have lower gas fractions than the m11s even by $z =3$. The m12s consume much more of their gas in star formation, and their gas fractions fall earlier in cosmic time, and more dramatically, than the m11s resulting in gas fractions of $f_{\rm gas} \approx 0.1-0.3$. The $z=0$ panels of the bottom section of Figure~\ref{fig:faceon_m11m12_gas_imshow} shows the disky morphology of \textbf{m12b} at $z=0$: we can clearly see spiral arms $z\sim0$, and the edge-on view shows a clear disk that was not evident at higher redshift. 

One of the most distinct features present in all eleven galaxies, across all redshifts, is the frequent stellar feedback. It is well-established that many FIRE galaxies are relatively bursty in star formation before the formation of disks \citep{Sparre2017, Faucher-Giguere2018, Orr2018, Orr2020, Orr2021, Stern2021, Gurvich2023, Hopkins2023, Sun2023}. Specifically, \citet{El-Badry2017},\citet{Angles-Alcazar2017a}, and \citet{Sparre2017} note that the resolved ISM and stellar feedback physics in FIRE-2 gives rise to `breathing modes' of star formation that often continue up until disk formation (z$\approx 0.4-0.7$ for MW-mass galaxies and \textbf{m11h}).

\section{Results}\label{sec:results}

\subsection{Mass and Energy Loadings}\label{subsec:overallfluxes}

As previously mentioned, loading factors are useful for the study of galactic outflows due to their ability to relate outgoing mass and energy with star formation. In line with convention, we calculate the mass-loading factor $\eta_M$ as

\begin{equation}
    \eta_M = \frac{\dot M_{out}}{\dot M_{\star}} \; ,
\end{equation}
and the energy-loading factor $\eta_E$ as
\begin{equation}
    \eta_E = \frac{\dot E_{out}}{\dot M_{\star}\cdot(E_{SN}/100 {\rm \; M_\odot})} \; ,
\end{equation}
where ($E_{SN}$/100 M$_\odot$) represents the mechanical energy injection rate per mass of stars formed, which is $10^{51}$ ergs per 100 M$_{\odot}$. Both loading factors are dimensionless measures of outflows, normalized by the contribution from star formation. 

Here, we use the 40 Myr-averaged SFR to calculate $\eta_M$ and $\eta_E$, as this SFR can be inferred from UV observations, and represents time scales long enough to trace injection of energy from core-collapse SNe from young clusters. We calculate values for the loading factors based on the overall outflows (gas with $v_{\rm out} > 0$).
We also compare the effects of imposing a velocity cut (three times the neutral gas velocity dispersion in the disk: $v_{cut} > 3\sigma_{\rm neut,z}$) on our calculated loading factors, which represent more conservatively estimated outflows, and present the results in Appendix~\ref{velocity_cuts}.

In presenting our loading factors, we compute these measures locally in the galaxies (per 750~pc pixels) and globally (averaged in each galaxy snapshot). Global values are divided by the area of the galaxy, approximated using the radius at which half of the SFR resides (cf. a galactic H$\alpha$ radius).

\subsubsection{Local Mass Loading}

\begin{figure*}
    \includegraphics[width=\textwidth]{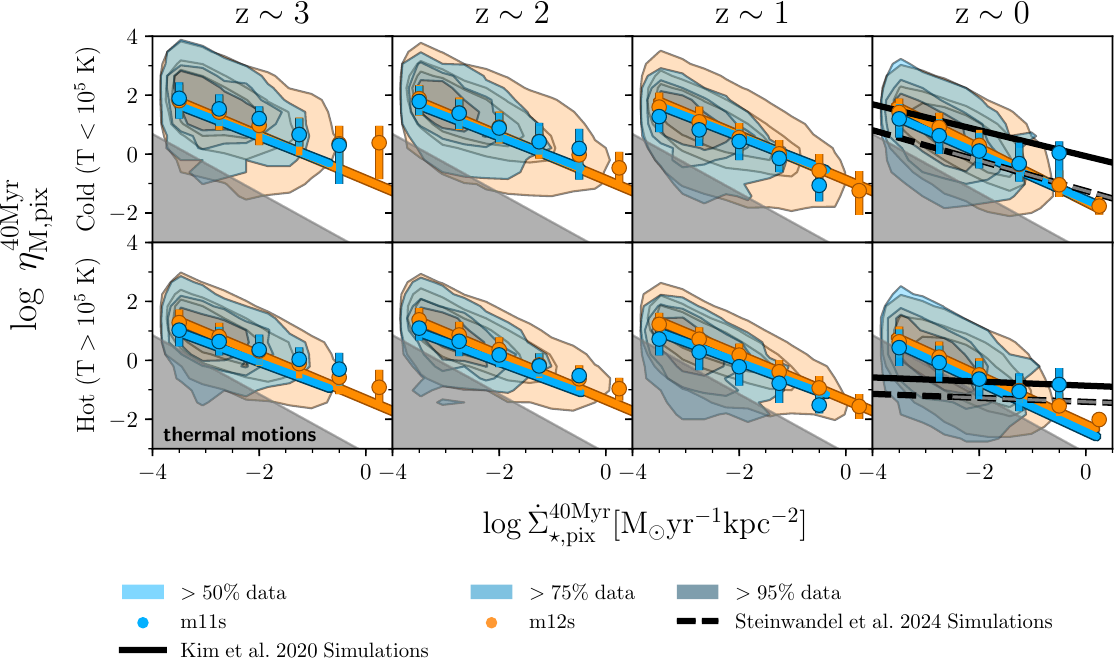}
    \caption{Star formation rate surface density (40 Myr average within the 750~pc pixels) vs. mass-loading factor ($\eta_{\rm M}$) for all eleven galaxies. Rows are separated by hot (T$>10^5$ K) and cold (T$<10^5$ K) gas phases.  Contours denote the 50\%, 75\%, and 95\% data inclusion regions, respectively, with blue contours belonging to the lower-mass galaxies (m11s) and orange to the MW-mass progenitors (m12s). Black solid lines are the best-fit mass loading factors calculated at one scale-height above the mid-plane in simulations by \citet{Kim2020a}, while dashed lines are the best-fit mass loading factors calculated at 1 kpc above the mid-plane in simulations by \citet{Steinwandel2024}; grey dashed lines represent an extrapolation of this fit. Points show the median $\eta_M$ and error bars represent the $\sim 1 \sigma$ error in 0.75 dex wide bins of SFR with at least fifty data-points. Shaded grey regions show areas where thermal motions of CGM particles may be dominating measurements.}
    \label{fig:etaM_tigress}
\end{figure*}

\begin{figure}
	\includegraphics[width=\columnwidth]{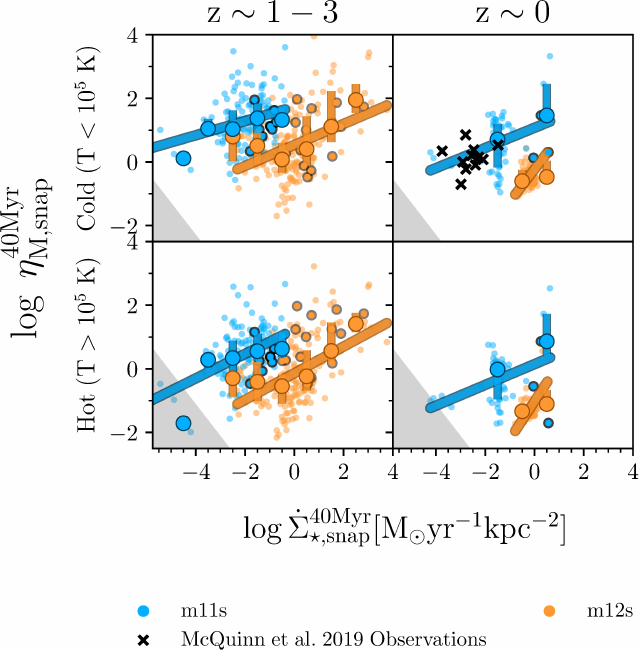}
    \caption{Star formation rate surface density and mass-loading factor similar to Figure~\ref{fig:etaM_tigress}, but calculated for the entire galaxy snapshot (i.e., globally averaged). Black x's denote observed low-mass galaxy outflows from \citet{McQuinn2019}. Points with black edges represent all superbubble snapshots from Appendix~\ref{eflux_SFR_all_gals}. Both galaxy groups now seem to occupy different regions of $\dot Sigma_{\star}$, with the m12 galaxies (orange) generally having higher $\dot \Sigma_{\star}$. Galaxies at high-redshift (left panels) have similar mass-loadings, while at $z\sim0$, the m11 SMC-mass galaxies have higher $\eta_M$.  Galaxy snapshots with the identified superbubbles appear to lie on the higher end of $\dot \Sigma_{\star}$. We find good agreement with the low-mass observed dataset.}
    \label{fig:etaM_global}
\end{figure}

Figure~\ref{fig:etaM_tigress} shows the spatially resolved distribution of $\eta_M$ with star formation rate surface density, with the top row showing the cold ($T<10^5$ K) gas and the bottom row the hot ($T>10^5$ K) gas phase. Each column represents the ten snapshots in each redshift range analyzed here ($z\sim 3-0$). Contours show the >50\%, >75\%, and >95\% data inclusion distributions of all pixels within the ten snapshots, while points represent the median $\eta_M$ value within 0.75 dex wide bins of SFR surface density (where each bin must have at least fifty pixels). Each median $\eta_M$ has corresponding 1$\sigma$ errorbars.  We also demarcate the region in which thermal CGM motions may dominate our measured `outflow' fluxes, see Section~\ref{thermalmotions} for details.

We show this distribution for both phases of the gas (hot and cold) across the redshifts, and also separate our galaxy populations into the m11s (blue) and m12s (orange). For each galaxy sample, we plot a best fit of the median $\eta_M$ values in each redshift panel. Equations for these lines of best fit can be found in Table~\ref{table:etaM_fits}.
We also plot theory predictions from \citet{Kim2020a} and \citet{Steinwandel2024}. 

Comparing the distributions of SMC-mass galaxies (m11s; blue) versus MW-mass progenitors (m12s; orange), we can see that at all times the two have very similar distributions of $\eta_M$; the only significant difference comes from the fact that the m12s host higher SFRs generally.   Both galaxy samples maintain identical slopes of $\eta_M$. The cold gas shows little signs of evolution with redshift, while the hot gas mass loadings show a subtle decrease at z$\sim$0 compared to z$\sim$1-3 (this is muddled somewhat by the change in the altitude at which the loading is measured).

Our spatially resolved cold gas mass-loadings are between the predictions from the highly resolved isolated galaxy simulations of \citet{Steinwandel2024}, and the tall-box {\scriptsize TIGRESS} simulations from \citet{Kim2020}. It appears that the cold gas in all of our FIRE-2 galaxies, at all redshifts, is carrying slightly more outflowing mass than the hot gas phase, consistent with other FIRE-2 measurements by \citet{Pandya2021}.
The spatially resolved hot gas $\eta_M$ is very similar to the cold gas in form with a similar power-law slope, though with an overall slightly lower normalization. This is in contrast with predictions of \citet{Kim2020} and \citet{Steinwandel2024}, who both predict a nearly flat relationship in the hot gas between $\eta_M$ and $\dot \Sigma_{\star}$ -- again similar to what we observe with the m11 galaxies.

\subsubsection{Global Mass Loading}

We also calculate $\eta_M$ globally, summing the outflows and star formation in the entire snapshot to calculate the loading factors.  
Figure~\ref{fig:etaM_global} presents the galaxy averaged mass loadings as Figure~\ref{fig:etaM_tigress}, and include observations from \citet{McQuinn2019} plotted as black x's in the upper-right panel, and simulation snapshot points with that include an identified superbubble outlined in black. Fit lines for global values of the loading factors can be found in Table~\ref{table:etaM_fits_global}.

With the global mass-loading values, we see an opposite slope as compared to the local calculations in Figure~\ref{fig:etaM_tigress}, and the two galaxy groups now occupy different spaces in $\dot \Sigma_{\star}$ and $\eta_M$. The SMC-mass m11 galaxies have lower star formation rate surface densities, as might be expected. And they also appear to have higher mass-loading values at all redshifts relative to their \textit{global} SFRs, speaking to the ability of winds escaping these lower-potential galaxies. We find good agreement with observations from \citet{McQuinn2019}, though the FIRE-2 m11 galaxies have higher SFRs as a sample. As with on the local scale, on the galaxy-scale mass-loading is dominated by the cold gas. The cold gas mass loading is $\sim$dex higher than the hot gas, across redshift. 

When it comes to the presence of superbubbles in these galaxy-averaged loading factors, superbubble snapshots tend to lie at the higher end of $\dot \Sigma_{\star}$, especially in the m11s (see lower-right panel of Figure~\ref{fig:etaM_global}). And the superbubbles in the m12s show higher mass-loadings in the hot gas at high redshift (we note that no significant superbubbles were identified in the m12 $z\approx0$ snapshots). 

\subsubsection{Local Energy Loading}

\begin{table}\caption{Fit lines for local values of $\eta_M$ (Figure~\ref{fig:etaM_tigress}) and $\eta_E$ (Figure~\ref{fig:etaE_tigress}), of the form $\log\eta = \gamma \log\dot\Sigma_\star^{\rm 40Myr} + \Delta$, where $\dot\Sigma_\star^{\rm 40 Myr}$ is in units of M$_{\odot}$ yr$^{-1}$ kpc$^{-2}$.
}
\label{table:etaM_fits}
    \begin{tabular}{lccccc} 
\hline
&  \multicolumn{2}{c}{z$\sim$1-3} & \multicolumn{2}{c}{z$\sim$0} \\
\hline
Loading Factor & $\gamma$ & $\Delta$ &  $\gamma$
& $\Delta$   \\ 
\hline
\hline
\multicolumn{5}{c}{\textbf{m11}} \\
$\eta_{M,\rm Hot}$
& -0.67 & -1.39 & -0.84 & -2.42   \\
$\eta_{M,\rm Cold}$
& -0.70 & -0.81 & -0.80 & -1.56 \\
$\eta_{M,\rm Hot \; vcut}$
& -0.67 & -1.23 & -0.82 & -2.00 \\
$\eta_{M,\rm Cold \; vcut}$
& -0.71 & -0.85 & -0.77 & -1.28 \\
\hline
$\eta_{E,\rm Hot}$
&-0.48  & -1.68 & -0.78 & -3.24 \\
$\eta_{E,\rm Cold}$
& -0.60 & -2.33 & -0.73 & -3.22 \\
$\eta_{E,\rm Hot \; vcut}$
& -0.52 & -1.46 & -0.78 & -2.71 \\
$\eta_{E,\rm Cold \; vcut}$
& -0.61 & -2.01 & -0.73 & -2.70 \\
\hline
\hline
\multicolumn{5}{c}{\textbf{m12}} \\
$\eta_{M,\rm Hot}$
& -0.76 & -1.34 & -0.83 & -2.18 \\
$\eta_{M,\rm Cold}$
& -0.76 & -0.88 & -0.91 & -1.62 \\
$\eta_{M,\rm Hot \; vcut}$
& -0.78 & -1.31 & -0.91 & -2.04 \\
$\eta_{M,\rm Cold \; vcut}$
& -0.76 & -0.86 & -0.88 & -1.42 \\
\hline
$\eta_{E,\rm Hot}$
& -0.61 & -1.41 & -0.78 & -2.63 \\
$\eta_{E,\rm Cold}$
& -0.68 & -2.00 & -0.90 & -2.53 \\
$\eta_{E,\rm Hot \; vcut}$
& -0.69 & -1.27 & -0.90 & -2.46 \\
$\eta_{E,\rm Cold \; vcut}$
& -0.73 & -1.68 & -0.88 & -2.26 \\
\hline
    \end{tabular}
\end{table}

\begin{figure*}
	\includegraphics[width=\textwidth]{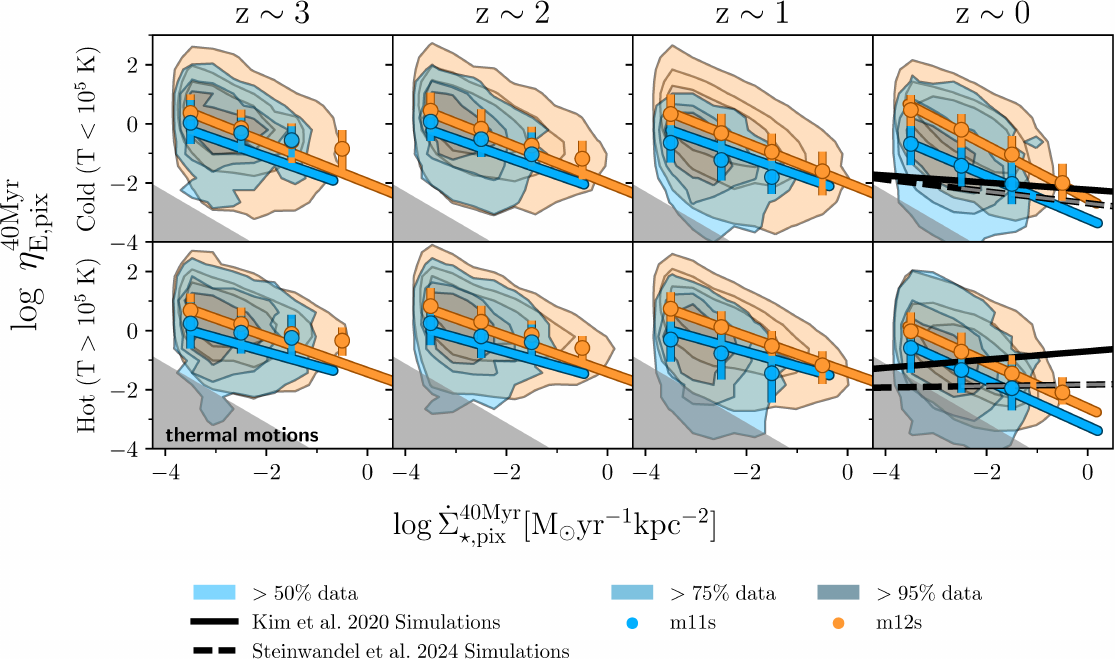}
    \caption{Star formation rate surface density (40 Myr average within the 750~pc pixels) vs. energy-loading factor ($\eta_{\rm E}$) for all eleven galaxies, in the style of Figure~\ref{fig:etaM_tigress}.  We generally find higher energy-loadings compared to {\scriptsize TIGRESS} and \citet{Steinwandel2024}, with a falling dependence in SFR.
    }
    \label{fig:etaE_tigress}
\end{figure*}

Figure~\ref{fig:etaE_tigress} displays the same information as Figure~\ref{fig:etaM_tigress}, with the exception that we now plot the local energy-loading factor, $\eta_E$, instead of $\eta_M$. As with $\eta_M$, the fit values for local $\eta_E$ can be found in Table~\ref{table:etaM_fits}.

The most contrast between the low-mass galaxies and MW-mass progenitors is here in the energy loadings. The MW-mass m12s, at all redshifts, have energy-loading factors that are nearly always larger than the m11s, reaching up to an order of magnitude difference in the cold gas at z$\approx$0 (see upper-right panel of Figure~\ref{fig:etaE_tigress}). This difference is not quite as large at higher redshifts. While it may be expected that low-mass galaxies (e.g., m11s) have higher energy loading factors due to their shallower potentials/more ``efficient'' outflows, we note that measuring outflows in the same way for both galaxy types can result in an inherent bias: larger galaxies like the m12s may need to have significantly more energy in their outflows in order to expel mass to the same relative height ($\rm \pm0.05 R_{vir}$ or 2H), due to their larger gravitational potentials. 

When comparing to \citet{Kim2020} and \citet{Steinwandel2024}, we find that our energy loadings in the cold gas generally are higher, though there is significantly less tension when considering the m11s. We find an opposite trend as compared to their simulations: falling energy loadings with SFR. One possible explanation for this is that when calculating loading factors, we use the 40Myr-averaged star formation rates. Some measurements of energy-loading may be biased due to this: any amount of detected `outflowing gas' that was launched by the tail end of a starburst $\sim$40 Myr ago could dramatically inflate loading factor values, whereas there is not a clear way in which energy loadings could be diluted to an extreme. This caveat may affect the resulting $\eta_E$ or $\eta_M$ distributions.

\subsubsection{Global Energy Loading}

\begin{figure}
	\includegraphics[width=\columnwidth]{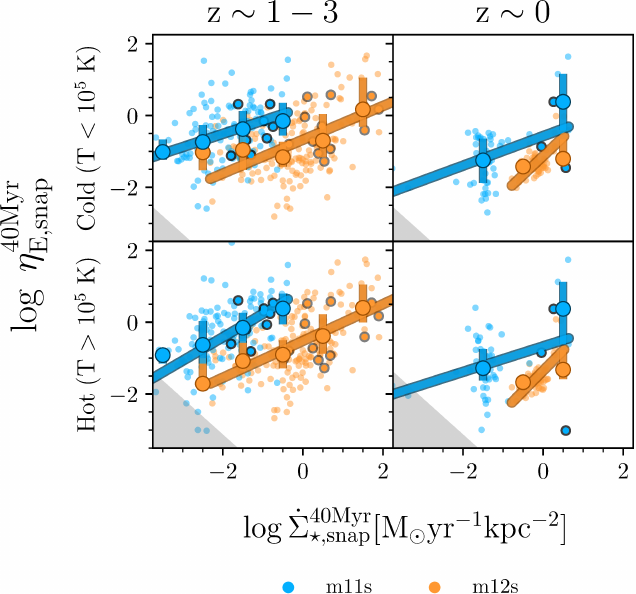}
    \caption{Star formation rate surface density (40 Myr average within the galaxy snapshot) vs. energy-loading factor ($\eta_{\rm E}$) for all eleven galaxies, in the style of Figure~\ref{fig:etaM_global}. Points with black edges represent all superbubble snapshots from Appendix~\ref{eflux_SFR_all_gals}. Galaxy snapshots with the identified superbubbles appear to lie on the higher end of $\dot \Sigma_{\star}$. Global galaxy distributions of $\eta_E$ are similar to $\eta_M$.}
    \label{fig:etaE_global}
\end{figure}

As with $\eta_M$, we also show the global (galaxy-averaged) distribution of the energy loading factors for this sample in Figure~\ref{fig:etaE_global}, with fits available in Table~\ref{fig:etaE_global}. 

Distributions in the global $\eta_E$ are very similar to those from Figure~\ref{fig:etaM_global}. The MW-mass progenitors appear to have higher global SFR surface densities than the low-mass m11s, and energy-loading factors at high-redshift are similar (albeit shifted in SFR). At high redshift, the energy loading is dominated by the hot gas, which matches theoretical predictions. At low redshift, however, the cold gas appears to have slightly more energy, especially in the m11s.  This may well be related, however, to the lack of significant large-scale outflows from the m12s, and generally smooth (in time) SFRs.

\begin{table}\caption{Fit lines for global values of $\eta_M$ (Figure~\ref{fig:etaM_global}) and $\eta_E$ (Figure~\ref{fig:etaE_global}), of the form $\log\eta = \gamma \log\dot\Sigma_\star^{\rm 40Myr} + \Delta$, where $\dot\Sigma_\star^{\rm 40 Myr}$ is in units of M$_{\odot}$ yr$^{-1}$ kpc$^{-2}$.
}
\label{table:etaM_fits_global}
    \begin{tabular}{lccccc} 
\hline
&  \multicolumn{2}{c}{z$\sim$1-3} & \multicolumn{2}{c}{z$\sim$0} \\
\hline
Loading Factor & $\gamma$ & $\Delta$ &  $\gamma$
& $\Delta$   \\ 
\hline
\hline
\multicolumn{5}{c}{\textbf{m11}} \\
$\eta_{M,\rm Hot}$
& 0.38 & 1.24 & 0.33 & 0.15 \\
$\eta_{M,\rm Cold}$
& 0.22 & 1.73 & 0.32 & 1.07  \\
$\eta_{M,\rm Hot \; vcut}$
& 0.37 & 1.16 & 0.29 & 0.04  \\
$\eta_{M,\rm Cold \; vcut}$
& 0.18 & 1.34 & 0.30 & 0.85  \\
\hline
$\eta_{E,\rm Hot}$
& 0.65 & 0.88 & 0.35 & -0.67  \\
$\eta_{E,\rm Cold}$
& 0.37 & 0.22 & 0.41 & -0.57 \\
$\eta_{E,\rm Hot \; vcut}$
& 0.63 & 0.93 & 0.32 & -0.71  \\
$\eta_{E,\rm Cold \; vcut}$
& 0.38 & 0.22 & 0.41 & -0.57  \\
\hline
\hline
\multicolumn{5}{c}{\textbf{m12}} \\
$\eta_{M,\rm Hot}$
& 0.42 & -0.14 & 1.19 &  -1.02 \\
$\eta_{M,\rm Cold}$
& 0.33 & 0.54 & 1.01 & -0.23  \\
$\eta_{M,\rm Hot \; vcut}$
& 0.42 & -0.22 & 1.19 & -1.02  \\
$\eta_{M,\rm Cold \; vcut}$
& 0.34 & 0.09 & 1.01 & -0.23  \\
\hline
$\eta_{E,\rm Hot}$
& 0.51 & -0.52 & 1.13 & -1.35 \\
$\eta_{E,\rm Cold}$
& 0.47 & -0.68 & 1.05 & -1.13 \\
$\eta_{E,\rm Hot \; vcut}$
& 0.52 & -0.43 & 1.11 &  -1.3 \\
$\eta_{E,\rm Cold \; vcut}$
& 0.51 & -0.78 & 1.01 & -1.22  \\
\hline

    \end{tabular}
\end{table}

\subsubsection{Contributions from Thermal Motions and/or CGM Gas}\label{thermalmotions}

To conservatively estimate the contribution of thermal motions in the inner CGM/halo of the simulations to our mass and energy flux measurements (essentially a resolution limit), we calculate here estimates for the mass and energy flux of a single gas element with representative temperatures and densities moving with thermal velocity across the flux measurement surface, yielding essentially minimum $\eta_M$ and $\eta_E$ values as a function of $\dot \Sigma_{\star}$ for the cool and hot gases.

Estimating that the mass and energy fluxes, for a single particle, given its smoothing length we have: $\dot M = h^2 \rho v$ and $\dot E = h^2 \rho v (v^2 + c_s^2/(\gamma-1))$, where we have taken the standard sound speed definition $c_s^2 = (\gamma -1) u$. And assuming that the velocity of the particle is the thermal sound speed, these fluxes simplify to: $\dot M = h^2 \rho c_s$ and $\dot E = h^2 \rho c_s^3 \gamma/(\gamma-1)$.  For fiducial values, we take the minimum baryonic particle mass in these FIRE-2 simulations, $m_{b,min}=7100 M_{\odot }$, and the velocity to be the sound speed of the gas (for the cool gas, $c_s \approx 12$ km/s assuming $\mu=0.59$ and $T=10^4$ K, while the hot gas has $c_s \approx 35$ km/s). Then, the smoothing length $h$ is found by assuming densities of $10^{-2}$ cm$^{-3}$ and $10^{-3}$ cm$^{-3}$ for the cool and hot gases, respectively, resulting in $h_{\rm cool}\approx 300$ pc and $h_{\rm hot}\approx 650$ pc.

The estimated `thermal' loading factors then scale by star formation surface density as:

\begin{equation*}
    \eta_{M, \rm cool}^{\rm thermal} \approx \frac{4.82 \times 10^{4}}{\dot \Sigma_{\star}} \hspace{5 mm} \& \hspace{5 mm} \eta_{E, \rm cool}^{\rm thermal} \approx \frac{4.60 \times 10^{7}}{\dot \Sigma_{\star}} \; , 
\end{equation*}
and
\begin{equation*}
    \eta_{M, \rm hot}^{\rm thermal} \approx \frac{7.18 \times 10^{4}}{ \dot \Sigma_{\star}}\hspace{5 mm} \& \hspace{5 mm} \eta_{E, \rm hot}^{\rm thermal} \approx \frac{6.88 \times 10^{6}}{\dot \Sigma_{\star}} \; , 
\end{equation*}

with $\dot \Sigma_{\star}$ in units of M$_{\odot}$ yr$^{-1}$ kpc$^{-2}$.

These `thermal' loading factors are visible as the shaded grey region in Figures~\ref{fig:etaM_tigress}--\ref{fig:etaE_global}. The grey shaded region essentially representing where our outflow measurements may be dominated by thermal motions of a single gas element at the simulation resolution scale in the inner CGM, as these particles are travelling at or somewhat below the sound speed.

Comparing this thermal limit with our spatially resolved distributions, we see that as much as half of the hot $\eta_M$ values at $z\approx 0$ (see lower-right panel of Figure~\ref{fig:etaM_tigress}) fall in this potentially unresolved thermal regime. The cool gas mass loadings are generally above this cutoff (i.e., it is highly supersonic), but at $z\approx0$ the low SFR end of the cold gas $\eta_M$ distribution abuts the thermal regime.

Concerning the spatially resolved energy loading factors $\eta_E$, the pixels are generally well above the thermal regime, though again the hot gas at $z\approx 0$ has some significant overlap.  This highlights the importance of resolving this diffuse phase of galactic winds/the CGM, especially in simulations with weak winds (like these FIRE m12 runs at $z\approx 0$). We show in Appendix~\ref{velocity_cuts} that our velocity cut of $3\sigma_{z}$ does not significantly help to separate our loading distributions any further from this `thermal' regime. This suggests that a (larger) outflow velocity cut does not allow us to any more cleanly interpret our results, rather that we appear to be at least marginally resolving the outflow dynamics in these simulations.

\subsection{Superbubbles Within the Simulations}\label{subsec:superbubbles}

\begin{figure*}
    \includegraphics[width=\textwidth]{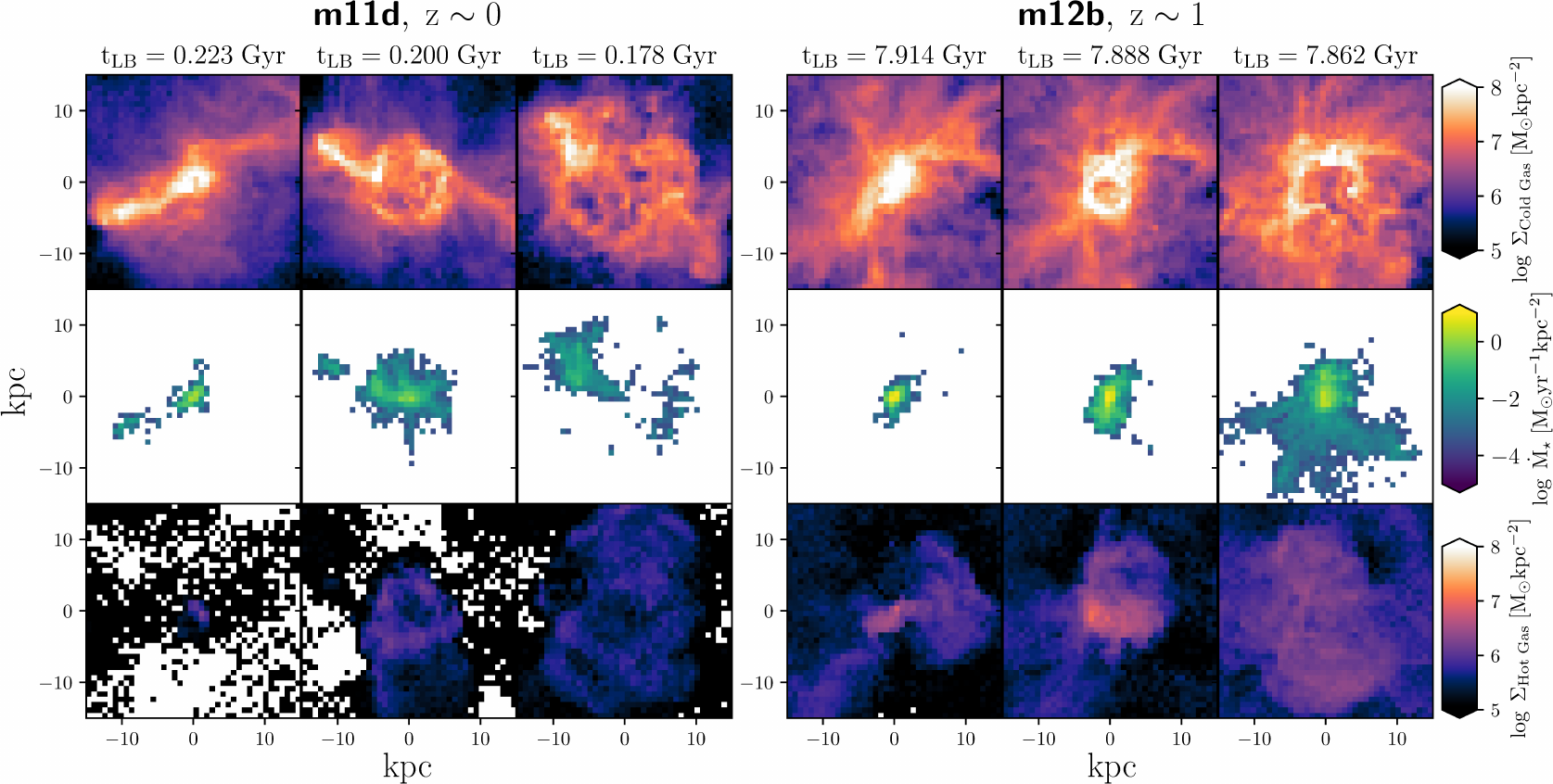}
    \caption{Face-on spatial distributions of two galaxies, \textbf{m11d} and \textbf{m12b}, where superbubble outbreak visibly occurs across consecutive snapshots. Left panels show low-mass galaxy \textbf{m11d} near z$\sim$0, while right panels show MW-mass progenitor \textbf{m12b} at z$\sim$1. Top row depicts the spatial distribution of the cold gas ($\rm T < 10^5$ K), middle row shows the 40Myr-averaged SFR, and the bottom row shows the hot gas ($\rm T > 10^5$ K.) In the first snapshot (first column) for each galaxy, star formation, cold gas, and hot gas can be seen to cluster in the central region of the galaxy. In the second snapshot (middle column), we see the formation of a bubble, i.e., a shell-like structure in the cold gas. Lastly (right columns), the diffuse hot gas is seen to break out from the bubble, thus concluding the evolution of the superbubble itself into the near-CGM.}
    \label{fig:m11_m12_bubble_imshow}
\end{figure*}

\subsubsection{Identifying Superbubbles}\label{subsubsec:identification}

We visually identified superbubbles within the simulations using the total gas distributions (see Figure~\ref{fig:faceon_m11m12_gas_imshow}) from all snapshots in our sample (oriented face-on). If we saw a clear bubble-like morphology in the gas, we then examined the snapshots directly before and after the `bubble' looking for additional tracers of superbubbles, similar to methods in \citet{Watkins2023a}. These include the `shell' of cold gas, as well as a diffuse hot gas component within the cold shell. In addition to these quantities, we used the 40 Myr-averaged SFR to trace whether an `active' star cluster was at the center of this cold gas shell (producing clustered SNe to drive the superbubble). If all of these conditions were met, we identified the event as a likely superbubble.

We conservatively identify a total of 9 primary superbubbles occurring throughout these eleven galaxies and four redshift ranges, with four of the superbubbles belonging to the m11 galaxies and five to the m12s. Certainly many more superbubbles occur between snapshots, or are not as clean-cut, so this is a lower limit to their occurrence rate in these simulations, and we highlight these as archetypal examples of this feedback mode. 
For the sake of conciseness, we present results in the main text for two specific superbubbles (the remainder can be found in Appendix~\ref{eflux_SFR_all_gals}): one in \textbf{m11d} at $z\sim0$, and one in \textbf{m12b} at $z\sim1$. These two particular superbubbles allow us to showcase results in one low-mass galaxy and one MW-mass progenitor, as well as at both low and high redshift. 

Figure~\ref{fig:m11_m12_bubble_imshow} shows the cold and hot gas surface densities, and the star formation rate surface density, for the two superbubble events of interest (left: \textbf{m11d}, right: \textbf{m12b}). Both galaxies are oriented `face-on', however, lacking a disk this distinction is not terribly important.
On the left side, \textbf{m11d}'s superbubble is shown, with the first column being the snapshot before the superbubble (about 22 Myrs beforehand), the middle column showing the first snapshot of the shell of cold gas expanding, and the third column as the snapshot immediately afterwards. The three columns together show a total span of about 44 Myrs in time from the first column's snapshot. In the top row, we have the surface density of the cold ($T<10^5$ K) gas, where the bubble itself is most visible. The middle row shows the 40 Myr-averaged star formation rate surface density, and the bottom row is the diffuse hot ($T>10^5$ K) gas. The right side of Figure~\ref{fig:m11_m12_bubble_imshow} shows the same information, but for galaxy \textbf{m12b}, spanning a total of about 50 Myrs in time. These panels of \textbf{m12b} depict a particularly stunning example of what one might expect from a superbubble: stellar feedback from the central star cluster couples together, creating the signature bubble shape in the cold gas, which then breaks open and fragments, releasing the hot gas component. The bubble in \textbf{m11d} appears to reach about 6 kpc in diameter before it fragments in the third snapshot, while the bubble in \textbf{m12b} is much smaller at 3-4 kpc in diameter.

\subsubsection{Superbubble Winds: Mass \& energy fluxes and the connection to star formation}\label{subsubsec:sbwinds_sf}

\begin{figure*}
    \includegraphics[width=0.9\textwidth]{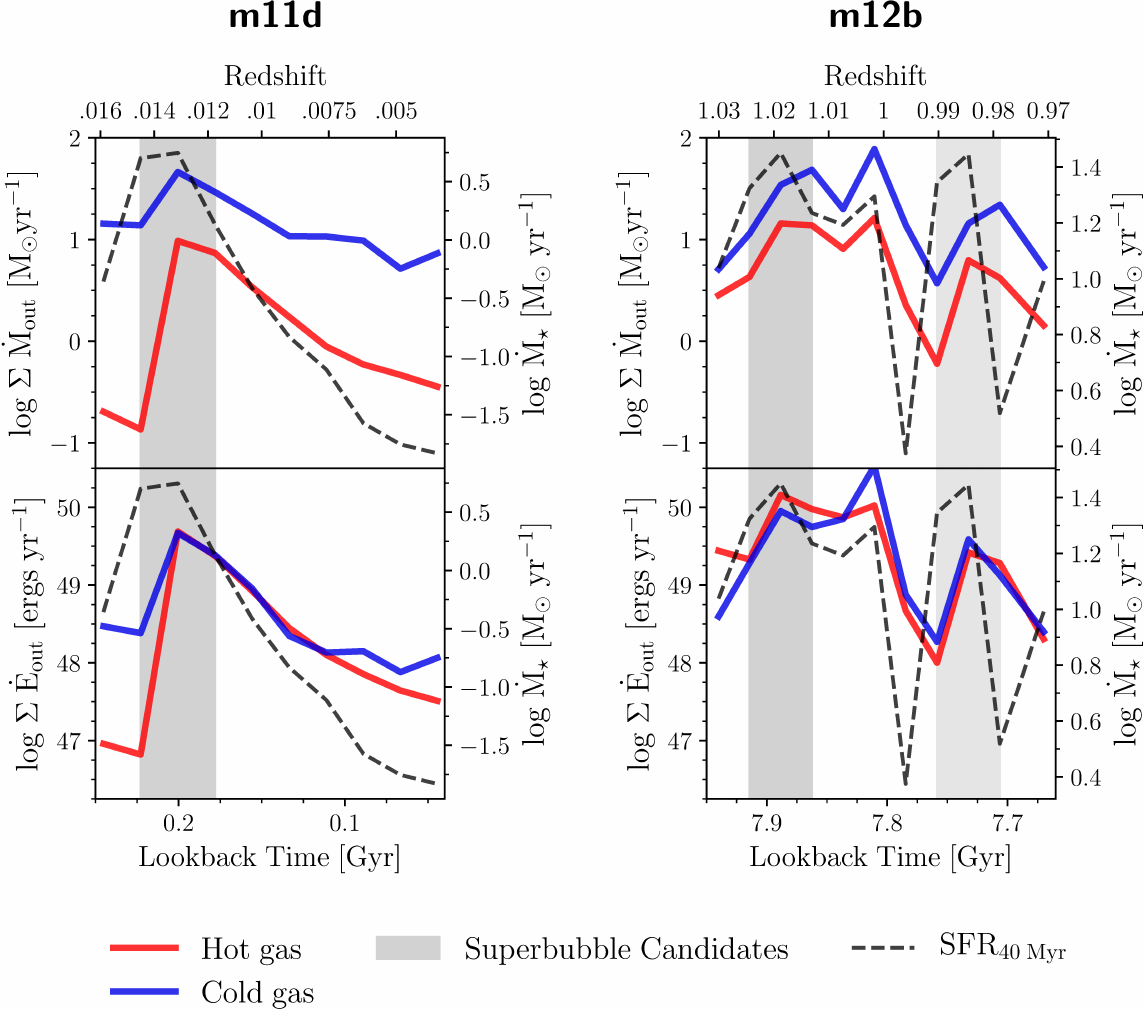}
    \caption{Flux out of the two galaxies shown in Figure~\ref{fig:m11_m12_bubble_imshow}, \textbf{m11d} \& \textbf{m12b}, for snapshots around the superbubbles. Top panels show the mass flux measured out of a surface 2H (\textbf{m11d}) and 0.05$\rm R_{vir}$ (\textbf{m12b}) above the galaxy center, respectively, where the surface height is determined by redshift. Bottom panels represent the energy flux out of the same galaxy surfaces. Lines are colored by gas temperature. The 40 Myr-averaged SFR is represented by the dashed black line. Shaded regions indicate snapshots where a large starbursting event is identified, with the primary bubbles shaded in dark grey (from Figure~\ref{fig:m11_m12_bubble_imshow}) and secondary events as lighter grey. The significant increase in cold gas mass flux corresponds with the cold gas ``cap'' of a superbubble expanding outwards.}
    \label{fig:m11_m12_flux}
\end{figure*}

Figure~\ref{fig:m11_m12_flux} depicts the total outflowing mass and energy flux from each galaxy in Figure~\ref{fig:m11_m12_bubble_imshow}, \textbf{m11d} and \textbf{m12b}, during the redshift range for which the bubble was identified. 
We plot both the hot and cold gas, and identify the time frame of the snapshots shown in Figure~\ref{fig:m11_m12_bubble_imshow} with a shaded dark grey region $\pm 1$ snapshot. Lighter shaded regions, if present, show smaller or secondary instances of stellar feedback.

In both  \textbf{m11d} and \textbf{m12b}, we see a sharp rise in the outflowing mass in all gas phases during the identified superbubble events. In \textbf{m11d} at redshift zero, during the snapshots of the superbubble, the total hot gas mass flux increases by more than an order of magnitude, while the cold gas increases about 0.5 dex. The outflows steadily decrease after the superbubble's fragmentation (right edge of grey shaded region). \textbf{m12b} shows a similar story: the hot and cold mass fluxes increase by 0.5-1.5 dex, then (on average) decreasing, though we do see another increase in the fluxes during this time span, suggesting a secondary event of strong stellar feedback.
For both \textbf{m11d} and \textbf{m12b}, we note that the cold phase of the gas carries a majority of the mass relative to the hot gas. 

As with the mass flux, we see an increase in the energy flux during the identified superbubble events, followed by a significant decrease. In \textbf{m11d}, the phase that seems to experience the most dramatic changes is the hot gas, rising up three orders of magnitude before decreasing, while in \textbf{m12b} the energy flux of both phases is almost equal. In \textbf{m12b}, we can see that the hot gas energy flux decreases by about 2 orders of magnitude some time after bubble fragmentation. In both events shown here, it is apparent that the outflowing energy flux of hot gas reaches a peak during the identified superbubble. 

In both galaxies, we see a significant increase in the star formation rate as the superbubble occurs, as much as an order of magnitude for \textbf{m11d}, followed almost immediately by a decrease after the bubble's fragmentation and release of hot and diffuse gas and energy. \textbf{m11d} shows a much clearer relationship, as the galaxy as a whole is experiencing less disruption at low $z$ than \textbf{m12b} near $z=1$. Tellingly, for every superbubble identified within the galaxies of this sample (see Appendix~\ref{eflux_SFR_all_gals}), we observe a direct relationship with the star formation rate (dashed black line) and the mass/energy fluxes.  Given the 40 Myr SFR averaging window, and the steep rise in SFRs with the bubble expansion, it is clear that the majority of star formation associated with the superbubble events occurs in $\lesssim 25$ Myr as the bubble is first expanding. This supports recent findings from JWST of a temporary increase in star formation during superbubbles in NGC 628 \citep{Barnes2023, Watkins2023a}, but it is not entirely clear that ``triggered star formation'' is a dominant scenario associated with superbubbles as opposed to a single large starburst. 

\subsection{What are the Conditions for Superbubble Break Out?}
\citet{Orr2022} presented a predictive model for whether a superbubble will  `break out' of the galactic disk (powered breakout; PBO), or stall within (powered stall; PS), based on parameters of the ISM, namely the local gas fraction $f_{\rm gas}$ and inverse dynamical time $\rm \Omega=v_c/R$. They then compare this theory with both simulations and observations.  \citet{Orr2022} makes several predictions regarding the balance of momentum injection into the ISM vs. CGM as a function of ISM properties, arguing that perhaps as much as $\sim60\%$ of SN momentum will go into the outflow, as opposed to the ISM \citep[with observational evidence corroborating this by][]{ReichardtChu2022}. None of the model assumptions \textit{require} a disk environment, and can therefore be tested by the range of galaxy morphologies/masses/sizes we use in this study \citep[see Section 7.4 of][]{Orr2022a}{}{}. 

\begin{figure*}
    \includegraphics[width=0.9\textwidth]{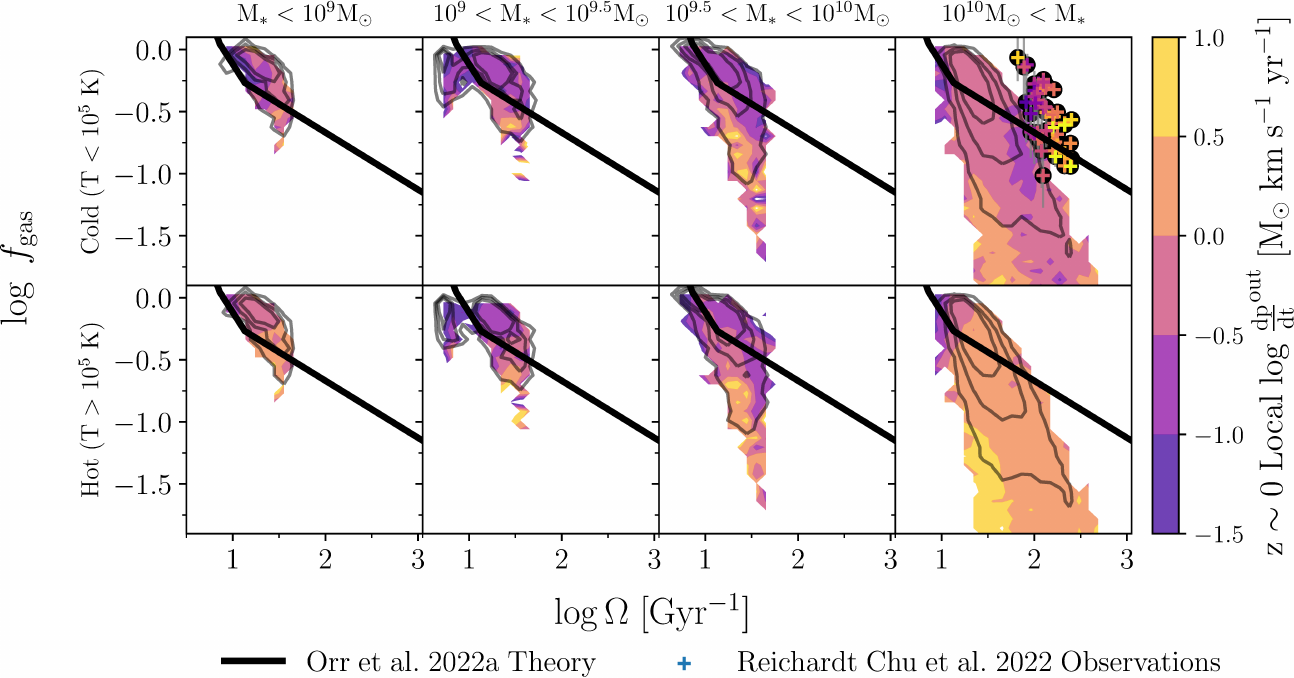}
    \caption{Gas fraction ($f_{\rm gas} =  M_{\rm gas}/(M_{\rm gas}+M_{\star})$), computed at the galaxy radius with 90\% of the stellar mass, vs.~ inverse dynamical time ($\Omega \equiv v_c/R$, with pixels weighted within each snapshot by $\dot\Sigma_\star^{\rm 40 Myr}$). Columns show bins of stellar mass. 2-D histogram and contours represent the spatially resolved data from all galaxies within the mass bin. 2-D histogram is colored by the local momentum flux at redshift $z=0$. Black line represents theoretical boundary in $f_{\rm gas}$-$\Omega$ derived in \citet{Orr2022}, above which superbubbles are expected to have sufficient momentum to break out of the ISM and drive outflows. Smaller points in the upper-right panel represent local universe cold gas outflow momentum flux observations by \citet{ReichardtChu2022}. Local momentum flux calculations at $z=0$ (contours) have comparable values to observations of \citet{ReichardtChu2022}, and show higher values in the hot gas than the cold gas }
    \label{fig:momentumflux_superbubble}
\end{figure*}

\begin{figure*}
    \includegraphics[width=0.9\textwidth]{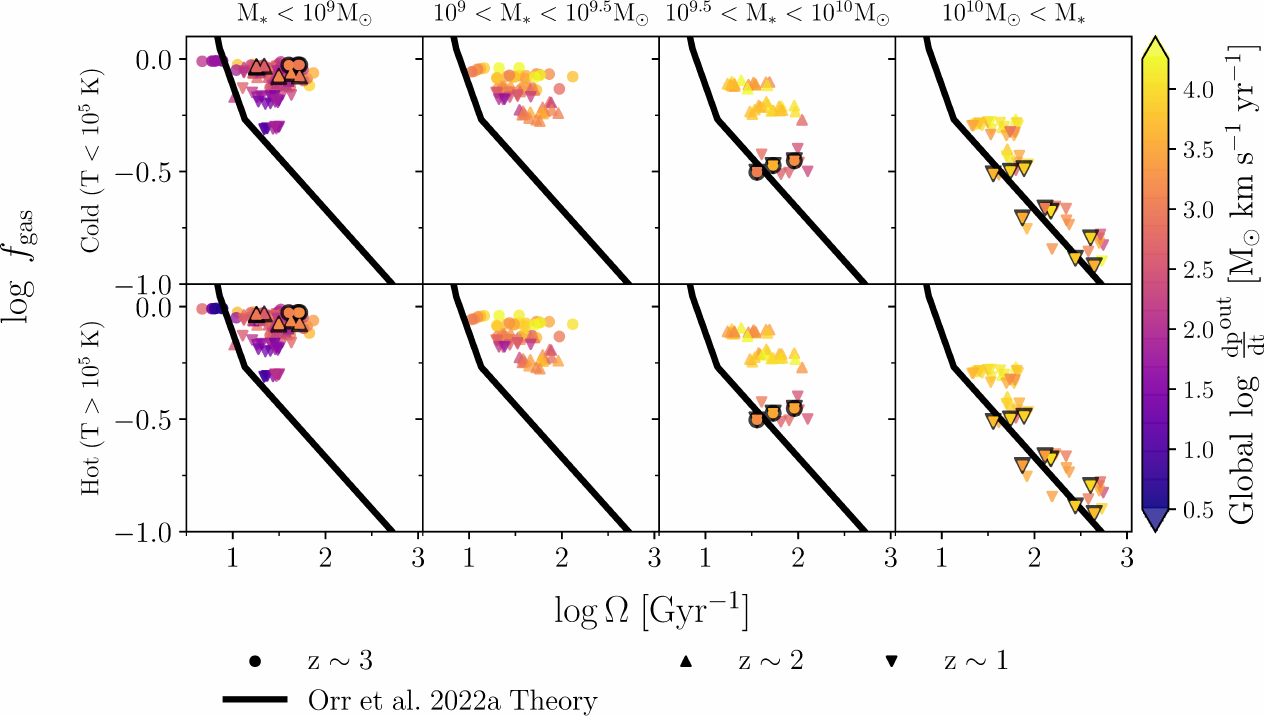}
    \caption{Same as Figure~\ref{fig:momentumflux_superbubble}, but for each galaxy snapshot in $z\sim1-3$. Snapshots with identified superbubbles in Appendix~\ref{eflux_SFR_all_gals} have a black background. Galaxy-averaged quantities (orange points) show the evolution of increasing gas fraction and decreasing inverse dynamical time with stellar mass/redshift, and increasing momentum flux. }
    \label{fig:momentumflux_superbubble_global}
\end{figure*}

Figure~\ref{fig:momentumflux_superbubble} shows our results for spatially-resolved local ($z\approx0$) momentum fluxes across our sample, plotted in the $\Omega$--$f_{\rm gas}$ phase space used by \citet{Orr2022a,Orr2022}. We do not distinguish between our SMC/LMC-mass galaxies (m11s) and MW-mass progenitors (m12s) here, instead organizing our results into bins of stellar mass. 
The shaded region shows the distribution of spatially resolved (750-pc scale) outflows at $z\approx 0$, colored by the average momentum flux value. A solid black line denotes the boundary line between powered superbubble breakout (PBO; right side of the boundary line) and powered stall (PS; left side of the boundary line) within the disk \citep[see Equation 9 of][]{Orr2022a}{}{}. For comparison, spatially resolved observations from a starbursting disk galaxy presented in \citet{ReichardtChu2022} are plotted as smaller points with black backgrounds in the upper-right panel of Figure~\ref{fig:momentumflux_superbubble} (cold gas and $M_\star >10^{10}$ M$_{\odot}$). 

The evolution in $f_{\rm gas}$ and $\Omega$ as stellar mass increases, with falling $f_{\rm gas}$ and increasing $\Omega$, is immediately evident. Quite similar momentum fluxes are seen in both the hot and cold gas for stellar masses $<10^{10}$ M$_\odot$.  We might interpret this as, for non-disk environments the winds are very mixed, and thus carry similar momenta.  However, for the $>10^{10}$ M$_\odot$ galaxies, the hot gas carries $\gtrsim$0.5 dex more momentum than the cold gas, suggestive of more ordered, hot winds being the general mode. 

Comparing our spatially resolved cold gas momentum fluxes with observations by \citet{ReichardtChu2022}, their values colored on the same scale in the Figure, in the range of about $10^{-1}-10^{0.50}$ $\rm M_{\odot} \; km \; s^{-1} \; yr^{-1}$, it is clear that their observed galaxy IRAS08 is significantly more gas rich than the FIRE-2 m12s in its center and hosts a far stronger outflow. However, there is agreement where our distributions overlap and generally a similar trend.  More extreme (in gas fraction, and presumably SFR) FIRE-2 disks would make for an excellent test of both the FIRE-2 model and the predictions of \citet{Orr2022}.

In Figure~\ref{fig:momentumflux_superbubble_global}, we show global (galaxy-averaged) outflow momentum fluxes in $f_{\rm gas}$--$\Omega$ space from $z\sim1-3$ snapshots. As with Figures~\ref{fig:etaM_global} \& ~\ref{fig:etaE_global}, we again mark the superbubble snapshots with black outlines.  Similar trends in gas fraction and dynamical time are seen, with decreasing $f_{gas}$ and increasing $\Omega$ with stellar mass and redshift. Momentum flux values clearly also increase with increasing stellar mass, moreso than with redshift alone. 

Most points lie in the powered breakout (PBO) region from \citet{Orr2022}, though we note that not every superbubble snapshot does. This may occur due to the `superbubble snapshots' including a snapshot before and after the bubble is first identified, and gas could be highly depleted following a superbubble disrupting the ISM. Furthermore, we notice no specific relationship between the superbubble snapshots and momentum flux; they do not have larger values as opposed to other points. However, at the highest stellar mass bins ($M_\star > 10^{9.5} {\rm \; M_\odot}$), they have among the lowest gas fractions.

\section{Discussion}\label{sec:disc}

It has been well-established that stellar feedback in the form of superbubbles drives some of the largest galactic-scale winds. However, the exact detailings and predominant physics involved in this process have been a topic of debate, particularly relating to the balance of regulating star formation in the ISM versus driving winds, the phase structure of the CGM, and various outflow properties.  

Both \citet{Kim2017a} and \citet{Fielding2018} highlighted the importance of spatial and temporal clustering of supernovae in driving superbubble winds, without which the supernova remnants would simply be expected to merge with the surrounding turbulent ISM.  The clustering, tied to the mass of the young star cluster, led to direct predictions by \citet{Fielding2018}
of loading factors for these winds. They predicted that energetic winds should have $\eta_M \approx 0.5-1$ at $z\sim2$, where the mass ejected by stellar feedback is on the order of the star formation rate, reminiscent of a `bathtub model' \citep{Dekel2014, Belfiore2019a}. These predictions are similar to what we see in our FIRE-2 galaxies at cosmic noon (evident in Figures~\ref{fig:etaM_tigress} \& \ref{fig:etaM_global}), particularly in the cold gas. 

Furthermore, \citet{Fielding2018} suggested that high mass-loading factors ($\eta_M \gg1$) in low-mass galaxies, similar to what we find at high redshift for the m11s, are often halo-scale quantities, as efficient venting of superbubble winds leads to less accretion onto the galaxy from the CGM, increasing the value of $\eta_M$ when averaged over scales comparable to the halo virial radius. Several studies have also suggested that hot outflows transport a majority of the energy from stellar feedback into the CGM \citep{Chevalier1985, Kim2020a, Li2017, Li2020, Fielding2022}. The argument follows that these hot energetic outflows deposit their heat in CGM, subsequently regulating star formation in the galaxy by preventing cold gas accretion \citep{Ostriker2010, Hopkins2014, Hayward2017, Li2020, Fielding2022}. 
By this logic, we might expect that galaxies with more efficient hot and energetic winds, such as low-mass galaxies with their shallower potential wells, will experience stronger star formation rate quenching following a large superbubble feedback event \citep{Muratov2015, Pandya2021}.  

We find this result in our simulations. As mentioned in Section~\ref{subsubsec:sbwinds_sf}, a clear correlation exists between the most energetic hot winds from a galaxy, usually occurring during a superbubble we have identified, and the 40 Myr-averaged star formation rate. Appendix~\ref{eflux_SFR_all_gals} contains a more complete picture of this result, showing our results for all the superbubbles we identify in this study. We note that the simulated galaxies, regardless of mass, experience a drop in star formation of about an order of magnitude following the superbubble's breakout and shell fragmentation, relative to their pre-superbubble SFRs. The drop in SFRs does appear to be more pronounced in the SMC-mass galaxies (see, e.g.,\textbf{m11d} and \textbf{m11e} in Figure~\ref{fig:all_m11_m12_Eflux_SFR}), regardless of redshift. And so, when the gravitational potential is lower winds appear to be far more effective at escaping the galaxy and suppressing star formation in FIRE-2 galaxies. 

Relatedly, \citet{Orr2022a}, using observational data, generally substantiated the suggestion by \citet{Orr2022} that superbubble feedback at higher redshifts ($z\geq1$) regulate the ISM differently: they predicted that superbubbles at $z\sim2$ should always drive true outflows and fountains, injecting most of the SN energy/momentum into the CGM. Our results here show a clear evolution in gas fraction and orbital time over redshift - this is largely linked to the corresponding evolution in stellar mass (see Figure~\ref{fig:momentumflux_superbubble}) - overall finding that gas fraction increases with increasing redshift, while the orbital time decreases. Nearly all of our galaxy snapshots, at all redshifts, lie within the `powered breakout' region from \citet{Orr2022}, indicating that these superbubbles should indeed drive galactic fountains and outflows if they occur. Though when we spatially resolve the inner regions of the m12s ($M_\star > 10^{10}$ M$_\odot$), \citet{Orr2022} would predict that those regions should not host significant, powered superbubble breakouts.  To wit, this aligns with our findings.  Whereas, for generally all of the spatially resolved regions of all of the lower mass galaxies, \citet{Orr2022} predicts superbubble breakout, i.e., none of the lower-mass galaxies should be expected to contain their supernova feedback, just as we observe in our study.

\subsection{Comparison to other Simulations}

Throughout the paper we have made comparisons with results from both the tall-box {\scriptsize TIGRESS} simulations \citep{Kim2020a}, and those of \citet{Steinwandel2024}.  Their works both study supernova-driven outflows in extremely high resolution, explicitly modeling a self-consistent ISM, feedback, and star formation at small scales: 2-8 pc in spatial resolution in a 500 pc tall-box disk patch in {\scriptsize TIGRESS}, and at $\sim $4 M$_{\odot}$, $\sim$1 pc resolution in an isolated LMC-mass galaxy in \citet{Steinwandel2024}. However, we caution the reader against too close a comparison with these simulations as the simulation methods are vastly different, with neither of their simulations including a realistic CGM into which the outflows are launched. In addition, the number of caveats in calculations can make comparison difficult, as no `gold standard' method exists for calculating outflows. We use slightly different temperature cuts on the gas, spatial resolution (750 pc), a 7100 M$_{\odot}$ mass resolution, and use a more diverse sample of galaxy types and redshifts.

Our spatially resolved $\eta_M$ values for cold gas in Figure~\ref{fig:etaM_tigress} fall between the of mass-loading values predicted by \citet{Kim2020a} and \citet{Steinwandel2024}. Interestingly, the FIRE-2 galaxies we analyzed have near identical dependencies for both gas phases on SFR, though with a lower normalization for the hot gas (it does not carry much of the mass). And although \citet{Steinwandel2024} and \citet{Kim2020a} use similar definitions for hot gas, we find a steeper dependence on SFR for mass loading of the hot phase, which may be related to our ability to resolve the hot phase on $\sim$kpc scale (motion at roughly the sound speed could dominate our hot gas fluxes). 

When considering the spatially resolved energy-loading factors, we find considerably more tension between our results and those of  \citet{Kim2020a} and \citet{Steinwandel2024}.  . Generally speaking we find significantly more energy in the cold gas phase, except for winds launched from the most star-forming of regions (top panels in Figure~\ref{fig:etaE_tigress}), though the SMC-mass galaxies (m11s; blue in Figure~\ref{fig:etaE_tigress}) are closer in magnitude. In the hot gas, we see little differences compared to the cold gas leading to an entirely opposite slope in $\eta_E$ and $\dot \Sigma_\star$ from other simulations, though significantly overlapping in overall energy at the high SFR end of our energy-loading distributions. 

There are several reasons why we could be seeing such higher energy loading factors in FIRE-2. One aspect could be the particular numerical implementation of supernova of stellar feedback in FIRE, combined with the burstiness of the simulated galaxies, \citep{Sparre2017, Faucher-Giguere2018, Orr2018, Orr2020, Orr2021, Stern2021, Gurvich2023, Hopkins2023, Sun2023}. The simulations by \citet{Kim2020} and \citet{Steinwandel2024} also lack significant features including a realistic cosmological background and a pre-existing CGM.  Recent studies also suggest, that including cosmic rays can have a significant effect on measured galaxy outflows \citep{Hopkins2021, Chan2022}. For example, \citet{Hopkins2021} finds that not including cosmic rays in L$_\star$ galaxies around $z=1-2$, included in redshift range and galaxy mass studied here, caused most outflows to simply be recycling material, while including cosmic rays leads to stellar feedback driving drastically different winds. \citet{Chan2022} finds a similar result, focusing specifically on the context of superbubbles in low-redshift L$_\star$ galaxies.

Due to the aforementioned caveats, we stress that care should be taken when comparing our loading factors with other simulations. There clearly exists a significant need to determine the extent to which physics (e.g., cooling, star formation, turbulence, cosmic rays), simulation parameters and methods (e.g., cosmological background, simulation scale, existing CGM, resolution), and outflow calculations (e.g., shells versus surfaces, velocity cuts, differing heights above the galactic center) affect loading factors and other observed properties of outflows driven by stellar feedback. 

\subsubsection{Within the FIRE Simulations}

\citet{Muratov2015}, using the original FIRE suite, found that galactic wind efficiency was dependent on halo mass, resulting in SMC/LMC-mass galaxies being extremely efficient at shutting off star formation through the expulsion of material in galactic winds. These results are aligned with what we find in the FIRE-2 galaxies, where more massive galaxies (especially those at high redshift) have less efficient galactic outflows compared to star formation on global scales \citep{Bassini2023}.
FIRE low-mass galaxies maintain their bursty star formation and wind recycling up through local times \citep{Muratov2015, Angles-Alcazar2017a, Hayward2017}, agreeing with the results we have found here.

\citet{Hopkins2023} provided a framework for understanding breathing modes of star formation/galactic winds: bursty star formation transitions to smoother star formation when the escape velocity becomes large enough that stellar feedback does not drive the escape of cold gas mass-loaded winds (i.e., the gravitational potential becomes deep enough to confine supernova ejecta). Stellar feedback then is largely confined to the disk and star formation becomes more steady. Additionally, \citet{Stern2021} and \citet{Hafen2022} showed that as the galaxy halo grows, the inner CGM virializes (becoming hot, promoting sub-sonic gas motion) and also promoting the confinement of outflows.

Another work using the FIRE-2 simulations, \citet{Pandya2021} found that the hot phase of gas carries both more mass and energy for MW-mass galaxies. This stands in contrast with other simulation predictions that the cold phase carries most of the mass and the hot phase carries most of the energy \citep{Kim2020a, Fielding2022}. In our work we find that on the local scale, the cold phase dominates the mass for the m12 galaxies, while the hot phase dominates the energy at high $z$, while the cold phase carries more energy at low $z$ (see Table~\ref{table:etaM_fits}). We see the same relationship for galaxy-averaged quantities. However, we note that we define outflows differently from \citet{Pandya2021}, who used surfaces instead of spherical shells further out at 0.1$\rm R_{vir}$, with different velocity cuts and averaged their loadings over Gyr-timescales. 

\subsection{Comparison to Recent Observations}

In Figure~\ref{fig:etaM_global} we compare directly with deep H$\alpha$ observations of outflows in six starbursting galaxies by \citet{McQuinn2019}. They suggest that true outflowing feedback-driven winds are preferentially identified in low-mass galaxies with centrally-concentrated star formation, similar to the findings by \citet{Fielding2018} and \citet{Orr2022a}, who predict feedback from high gas surface density central regions produce winds more likely to escape the galaxy. Indeed, we identified such winds in the SMC-mass galaxy \textbf{m11d}, that does host centrally clustered star formation (see the center panels of Figure~\ref{fig:m11_m12_bubble_imshow}).

We also compare the $\sim$kpc-scale momentum flux in the outflows of our Milky Way mass galaxies with similarly spatially resolved observations of IRAS08, a starbursting disk galaxy, by \citep{ReichardtChu2022}, as described in Section~\ref{subsec:superbubbles}. Though we measure outflowing gas in all of our m12s, it is clear to see that the FIRE-2 disk galaxies generally do not form massive nuclear star clusters capable of hosting fast supernova-driven winds, the likes of which are found in IRAS08 or, e.g., M82 \citep{Leroy2015b}.  Most of the spatially resolved `winds' in the Milky Way mass galaxies ($M_\star \sim 10^{10}$ M$\odot$) originate from the galactic outskirts (see Figure~\ref{fig:momentumflux_superbubble}), though where there is overlap with the distribution of momentum fluxes seen in IRAS08, we find general agreement in their magnitude.  Feedback models for regulating the ISM have highlighted the trade-off between driving turbulence in the ISM and galactic outflows \citep{Ostriker2010, Faucher-Giguere2013, Hayward2017, Krumholz2018, Orr2022}. Our results, showing a lack of strong wind driving from the central regions, are in agreement with the superbubble breakout prediction of \citet{Orr2022}.

Recent studies from ALMA and JWST have also directly helped constrain theoretical predictions of superbubble feedback. Work with JWST in NGC628 by \citet{Mayya2023} uncovered the temporally extended star formation episode driving a superbubble shell (the $\sim$kpc hole in the outer disk of that galaxy), in-line with our results in Figure~\ref{fig:m11_m12_flux} finding that an increase in star formation rate persists (i.e., some evidence of `positive feedback') during the galaxy's identified superbubble episode. This result is likely capturing the continued stellar feedback that contributes to driving the bubble itself, capturing an ongoing pattern of stellar feedback and star formation. 

\section{Summary \& Conclusions}\label{sec:conc}

In this paper, we presented an analysis of supernova feedback-driven outflows in eleven FIRE-2 galaxies, six of which are MW-mass progenitors and five of which are SMC/LMC-mass progenitors across a redshift range from $z \approx 3-0$. At higher redshifts ($z\geq1$) we calculate outflows through a surface at a distance $\pm0.05R_{vir}$ above/below the galaxy, while at $z=0$ we calculate outflows at two galaxy-averaged gas scale heights above/below the galaxy. We note that both choices have similar values (see Appendix~\ref{flux_surface_changes}). 

Comparing our findings from the FIRE suite with other simulations and recent observations, we found good agreement with observed mass loadings of winds, but elevated energy loadings compared to the {\scriptsize TIGRESS} simulations, though we note the simulations differ in kind including star formation rates and cosmological context (their tallboxes lacking a CGM).  We picked out several remarkable instances of galaxy-wide superbubbles occurring in simulations, and characterized the behavior of their resulting outflows.

Our main findings can be summarized by the following points:

\begin{enumerate}
    \item On 750 pc scales, we find negatively sloped power laws in both spatially resolved mass loadings $\eta_M$ and energy loadings $\eta_E$ vs.~star formation rate surface density $\dot \Sigma_\star$, with the steepest slopes found at $z\approx0$. Our values for cold gas $\eta_M$ are comparable with observations from \citet{McQuinn2019} and simulations by \citet{Steinwandel2024} and \citep{Kim2020a}. We predict higher hot gas mass loadings across redshift than either the {\scriptsize TIGRESS} simulations or those of \citet{Steinwandel2024}. 
    
    \item On 750~pc scales, the MW-mass galaxy progenitors (m12s) and SMC-mass galaxies (m11s) have nearly identical mass loadings as a function of star formation rate surface density.  The MW-mass galaxies do have higher energy loadings $\eta_E$ on these scales, which may be due to the halo potentials; the galaxy must launch much more energetic outflows to reach the same defined height (2H, 0.05$R_{vir}$) as the smaller m11s.

    \item On the galaxy scale, the SMC-mass m11 galaxies have higher $\eta_M$ in both gas phases than the MW-mass progenitors (m12s), and the cold phase carries most of the mass at all redshifts. The m11s also appear to carry more energy. This may be simply the result that the galaxies, though launching similar mass/energy-loaded winds locally, have lower global SFRs. At high redshifts, the energy loading is dominated by the hot gas, while at low $z$ the energy distribution is closer to equipartition. 
    
    \item These simulated galaxies all exhibit bursty star formation modes prior to the formation of disks. Following starburst events, in nearly all cases, we see significant outflowing mass and energy flux originating from clearly identifiable superbubble events.
    
    \item  All snapshots of all galaxies in our analysis have galaxy-averaged ISM properties that correspond to the `Powered Breakout' regime from \citet{Orr2022a}, with values that are comparable to observations from DUVET \citep{ReichardtChu2022}, suggestive of the idea that most if not all of these simulations host star-forming regions capable of driving outflows.

    \item  Variation in measured outflow rates, in both $\eta_M$ and $\eta_E$, can vary significantly (an order of magnitude of more) depending on where outflows are measured (see Figure~\ref{fig:etas_SFRsd_multipleheights}), and thermal motions from the inner CGM may play a significant role in these measurements.
\end{enumerate}

Quantifying the effects of clustered stellar feedback on galaxies of all types across cosmic time is essential to grasping what modes of star formation are capable of driving the strong galactic outflows seen in the universe. We know these outflows have direct consequences on galaxy formation: affecting accretion, star formation, the mass-metallicity relation, \textit{etc.} and a firm understanding of these processes directly constrains our theories of galaxy formation \& evolution. Future theoretical work understanding galaxy-scale superbubbles and feedback-driven winds will become only more necessary as observational constraints on mass and energy loadings at all redshifts tighten, and more detailed kinematics of superbubbles are measured.  

\section*{Acknowledgements}

We would like to thank Drummond Fielding for useful conversations and comments that greatly improved the manuscript.
We ran simulations using: XSEDE, supported by NSF grant ACI-1548562; Blue Waters, supported by the NSF; Frontera allocations AST21010 and AST20016, supported by the NSF and TACC; Pleiades, via the NASA HEC program through the NAS Division at Ames Research Center. L.E.P. is thankful to the Flatiron Institute for continued financial support and resources.  The Flatiron Institute is supported by the Simons Foundation.  BB is grateful for generous support by the David and Lucile Packard Foundation and Alfred P. Sloan Foundation. AW received support from: NSF via CAREER award AST-2045928 and grant AST-2107772; NASA ATP grant 80NSSC20K0513.  CAFG was supported by NSF through grants AST-2108230, AST-2307327, and CAREER award AST-1652522; by NASA through grants 17-ATP17-0067 and 21-ATP21-0036; and by STScI through grants HST-GO-16730.016-A and JWST-AR-03252.001-A.

%%%%%%%%%%%%%%%%%%%%%%%%%%%%%%%%%%%%%%%%%%%%%%%%%%
\section*{Data Availability}

The data supporting the plots within this article are available on reasonable request to the corresponding author. A public version of the GIZMO code is available at \url{http://www.tapir.caltech.edu/~phopkins/Site/GIZMO.html}. Additional data including simulation snapshots, initial conditions, and derived data products are available at \url{https://fire.northwestern.edu/data/}. The FIRE-2 simulations are publicly available \citep{Wetzel2023} at \url{http://flathub.flatironinstitute.org/fire}.



\bibliographystyle{mnras}
\bibliography{ref} % if your bibtex file is called example.bib

%%%%%%%%%%%%%%%%%%%%%%%%%%%%%%%%%%%%%%%%%%%%%%%%%%

%%%%%%%%%%%%%%%%% APPENDICES %%%%%%%%%%%%%%%%%%%%%

\appendix

\section{Impact of a Minimum Velocity Cut on Measured Outflows} \label{velocity_cuts}

\begin{figure*}
	\includegraphics[width=0.9\textwidth]{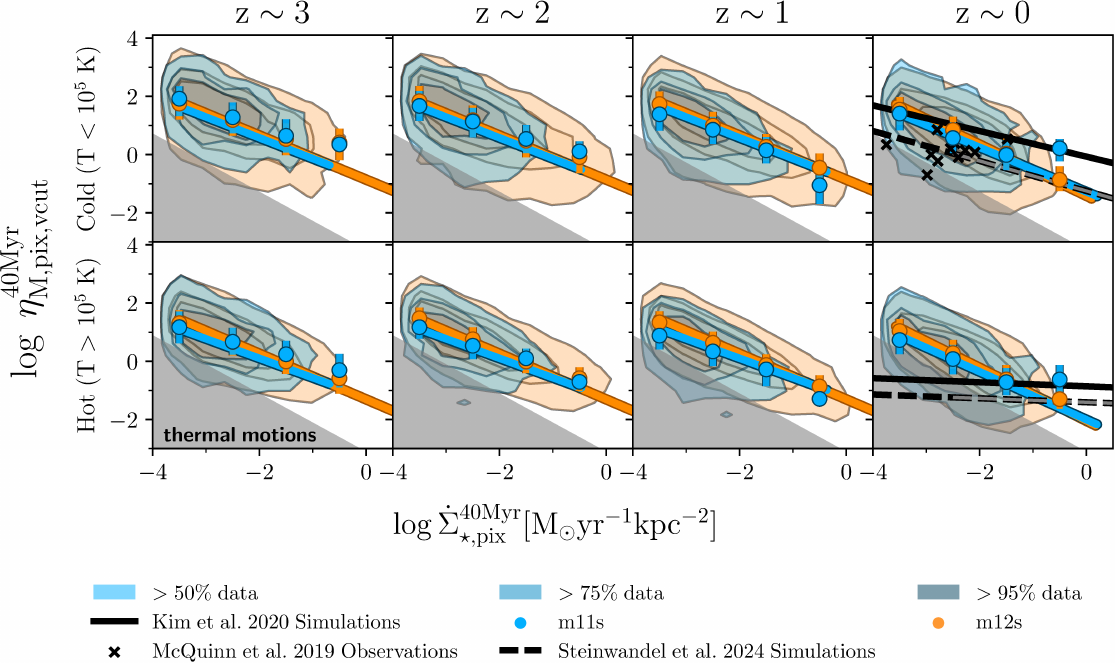}
    \caption{Same quantities and style as Figure~\ref{fig:etaM_tigress}, but now requiring that outflows have a velocity out of disk of at least three times the neutral gas velocity dispersion (v $> 3\sigma_z$). We see little differences in median mass-loading factors calculated with and without the velocity cut, though somewhat reduced scatter in the distribution.}
    \label{fig:etaM_velocitycut}
\end{figure*}

\begin{figure*}
	\includegraphics[width=0.9\textwidth]{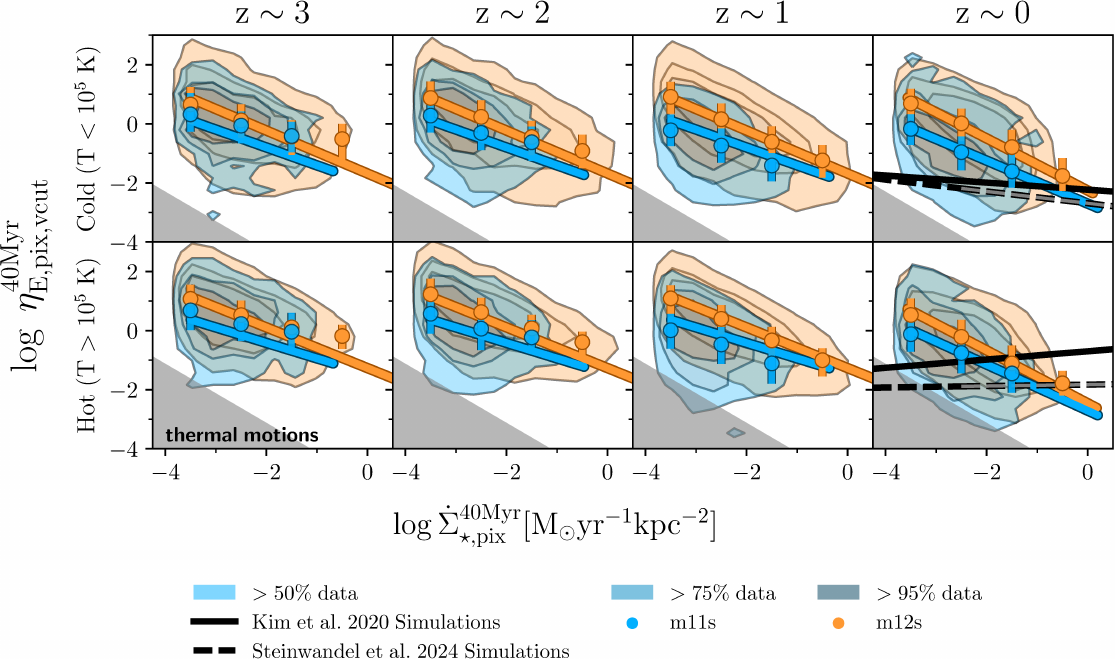}
    \caption{Same quantities and style as Figure~\ref{fig:etaE_tigress}, but now requiring that outflows have a velocity out of disk of at least three times the neutral gas velocity dispersion (v $> 3\sigma_z$). We see little differences in the median energy-loading factors with and without the velocity cut in place, though somehwat reduced scatter in the distribution.}
    \label{fig:etaE_velocitycut}
\end{figure*}

\begin{figure}
	\includegraphics[width=\columnwidth]{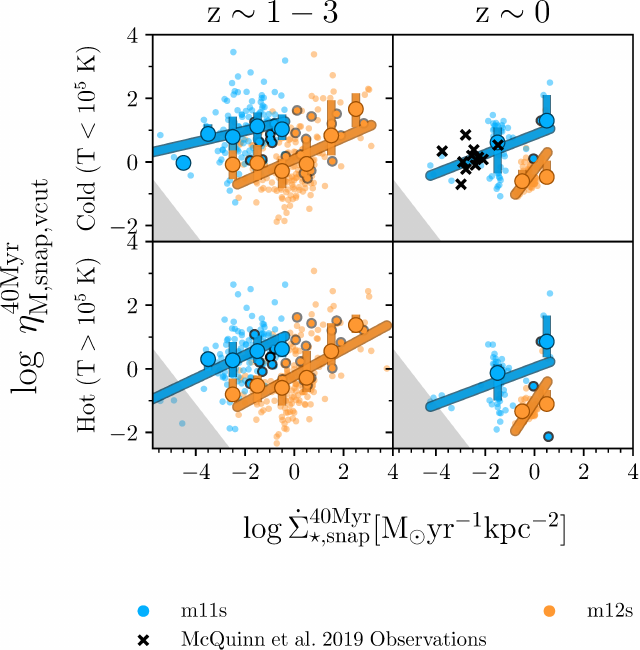}
    \caption{Same quantities and style as Figure~\ref{fig:etaM_global}, but now requiring that outflows have a velocity out of disk of at least three times the neutral gas velocity dispersion (v $> 3\sigma_z$). We see little differences in median mass-loading factors calculated with and without the velocity cut, though somewhat reduced scatter in the distribution.}
    \label{fig:etaM_global_velocitycut}
\end{figure}

\begin{figure}
	\includegraphics[width=\columnwidth]{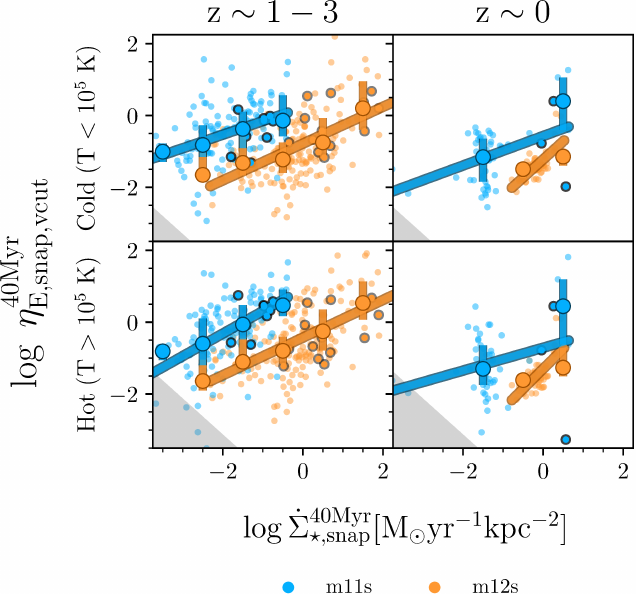}
    \caption{Same quantities and style as Figure~\ref{fig:etaE_global}, but now requiring that outflows have a velocity out of disk of at least three times the neutral gas velocity dispersion (v $> 3\sigma_z$). We see little differences in median mass-loading factors calculated with and without the velocity cut, though somewhat reduced scatter in the distribution.}
    \label{fig:etaE_global_velocitycut}
\end{figure}

In many studies of galactic winds, a minimum outflow velocity cut is often done. To ensure we are selecting \textit{outflows} (and not just the turbulent ISM), we also calculate mass and energy loading factors imposing a velocity cut of three times the neutral gas velocity dispersion ($v_z > 3\sigma_z$). We present the results for spatially resolved loadings in Figures~\ref{fig:etaM_velocitycut} \& \ref{fig:etaE_velocitycut}, for $\eta_M$ and $\eta_E$, respectively. And the results for galaxy-averaged loadings are shown in Figures~\ref{fig:etaM_global_velocitycut} \& \ref{fig:etaE_global_velocitycut}, for $\eta_M$ and $\eta_E$, respectively.  This cut is about the maximal observed velocity of the cold component of a number of local superbubbles \citep{Orr2022a}. The values of the corresponding fits can also be found in Table~\ref{table:etaM_fits}. However, we find little differences in either loading factor or fluxes with the velocity cut imposed, implying that most outflows identified by our calculations are by-and-large already travelling above this velocity.

\section{Dependence of Outflow Fluxes on Height/Distance of Measured Flux Surface} \label{flux_surface_changes}

\begin{figure}
	\includegraphics[width=1\columnwidth]{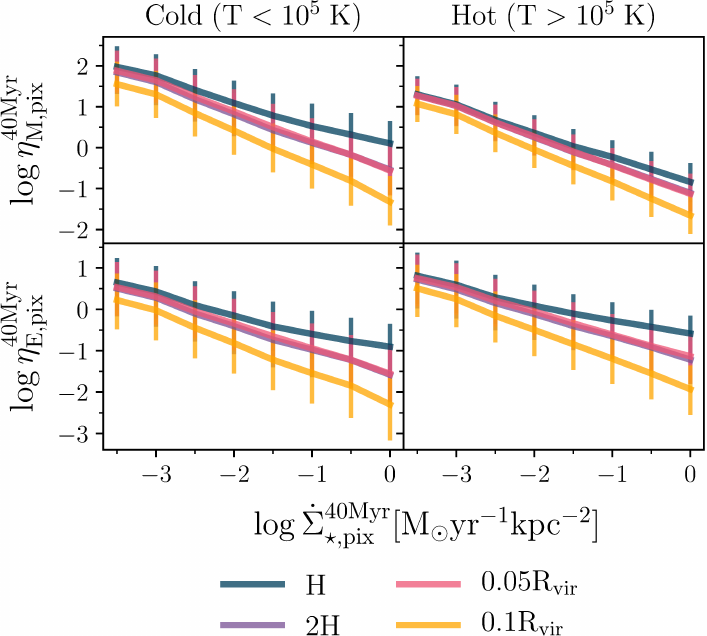}
    \caption{Star formation surface density versus mass- (top row) and energy-loading factors (bottom row) for all galaxies and snapshots, for multiple defined surface heights including: one (galaxy-averaged) gas scale height $H$, $2H$, 0.05 $ R_{\rm vir}$, and 0.1 $R_{\rm vir}$.  Errorbars denote the first and third quartile of the loading factor, while lines are medians for each bin in SFR surface density (0.5 dex wide). Generally, fluxes are lower as the surface through which they are measured moves away from the galaxy.  The mass-loading of hot gas is least affected by this, whereas the energy-loading of the cold gas is most strongly affected.}
    \label{fig:etas_SFRsd_multipleheights}
\end{figure}

In Figure~\ref{fig:etas_SFRsd_multipleheights}, we show how the various loading factors change if we choose to place the surface through which outflows are defined at different heights above the galaxy. We do not distinguish between galaxy type or redshift here, or impose a cut on velocity at the measured surface, instead focusing solely on the differences in height. 

We can see that there is little difference in flux between any of the heights ($H$, $2H$, $\rm 0.05R_{vir}$, $0.1R_{vir}$) for the hot gas mass-loading factor (upper-right panel). Other quantities, unsurprisingly however, shows a difference of up to an order of magnitude between the lowest surface height, $H$, and the highest surface height, $\rm 0.1R_{vir}$, in a consistent manner: there is more flux through surfaces closer to the galaxy mid-plane. We include this comparison here to show to what extent our fiducial chosen heights affects our results. 

While we have shown that the relationship between loading factors and the star formation rate surface density are affected by the chosen surface height, we stress that we frequently analyze how the gas and energy flux differ over a period of time where incidents of strong stellar feedback have been identified (see, e.g., Figure~\ref{fig:m11_m12_flux}).  It becomes important in these cases that the surface where we define outflows is not only at an appropriate height compared to the galaxy as a whole, but that this surface height is \textit{consistent}. Otherwise, our measurements of outflow properties may not reflect the true effect of feedback. For example, if we define the measured outflow surface at the gas scale height, when a single massive superbubble causes gas to be disrupted on a galaxy-wide scale, the gas scale height can dramatically change from snapshot to snapshot. If we measure outflow properties through the surface one or two scale heights above the galaxy during one of these events, the outflow properties might just reflect the growing/expanding `cold cap' of the superbubble as the galaxy quite literally breaks apart. In contrast, measurements at fractions of the virial radius are much more stable against large-scale disruptions as the virial radius grows smoothly with cosmic time.

\section{Energy Fluxes and Star Formation Rates for Individually Identified Superbubble Events} \label{eflux_SFR_all_gals}

\begin{figure*}
	\includegraphics[width=\textwidth]{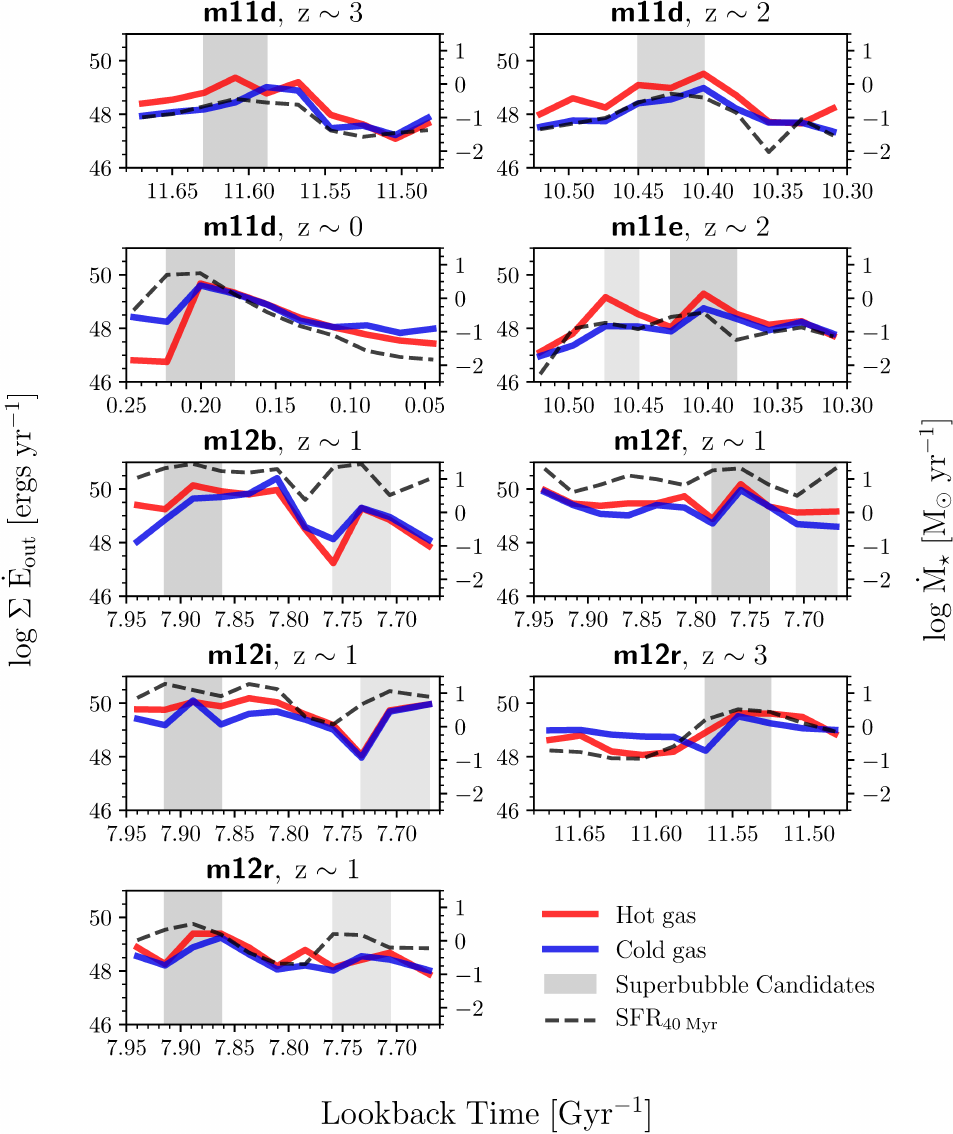}
    \caption{Hot energy flux out of each low-mass galaxy in this sample for which a superbubble was visually identified between snapshots, in the style of Figure~\ref{fig:m11_m12_flux}. In nearly all cases, hot energy flux peaks during the superbubble, and corresponds to a significant increase in SFR, which increases in some cases by as much as an order of magnitude. For each galaxy here, regardless of redshift, the SFR decreases after the superbubble.}
    \label{fig:all_m11_m12_Eflux_SFR}
\end{figure*}

Figure~\ref{fig:all_m11_m12_Eflux_SFR} shows the evolution in the energy flux of the hot gas component as presented in Figure~\ref{fig:m11_m12_flux}, but for all identified superbubbles in all individual galaxies of this study. 

These events occur in all four redshift bins ($z\sim3-0$) and both galaxy types of this study. In addition, we note that the same relationship is observed in every single identified superbubble: the star formation rate reaches a peak (typically for the entire time plotted) during the superbubble snapshots, then quickly drops. This drop in star formation is between 0.5-1 dex, with the exception of \textbf{m12b}, which experiences a smaller drop of 0.25-0.5 dex. 

%%%%%%%%%%%%%%%%%%%%%%%%%%%%%%%%%%%%%%%%%%%%%%%%%%

% Don't change these lines
\bsp	% typesetting comment
\label{lastpage}
\end{document}